\documentclass[aps,prb,amssymb,showpacs,twocolumn,longbibliography]{revtex4-2}

\usepackage{graphics,graphicx,amsmath}
\usepackage{tabularx,float}
\usepackage{epsfig}
\usepackage{subfigure}
\usepackage{bm}
\usepackage{stmaryrd}
\usepackage{mathtools}
\usepackage{wasysym}
\usepackage{bbm}
\usepackage[T1]{fontenc}
\usepackage[french,english]{babel}

\usepackage{mathrsfs} 
\usepackage{physics}

\usepackage{color}
\usepackage{verbatim}
\graphicspath{{fig/}}

\DeclareMathOperator{\ad}{ad}
\DeclareMathOperator{\sgn}{sgn}
\DeclareMathOperator{\Cov}{Cov}

\begin{document}

\date{\today}

\title{Normal state quantum geometry, non-locality and superconductivity}

\author{Florian Simon}
\affiliation{Laboratoire de Physique des Solides, Universit\'e Paris Saclay,
CNRS UMR 8502, F-91405 Orsay Cedex, France}

\begin{abstract}
We investigate aspects of the relation between the quantum geometry of the normal state (NS) and the superconducting phase, through the lens of non-locality. By relating band theory to quantum estimation theory, we derive a direct momentum-dependent relation between quantum geometry and the quantum fluctuations of the position operator. We then investigate two effects of the NS quantum geometry on superconductivity. On the one hand, we present a physical interpretation of the conventional and geometric contributions to the superfluid weight in terms of two different movements of the normal state charge carriers forming the Cooper pairs. The first contribution stems from their center-of-mass motion while the second stems from their zero-point motion, thereby explaining its persistence in flat-band systems. On the other hand, we phenomenologically derive an emergent Darwin term driven by the NS quantum metric. We show its form in one and two-body problems, derive the effective pairing potential in $s$-wave superconductors, and explicit its form in the case of two-dimensional massive Dirac fermions. We thus show that the NS quantum metric screens the pairing interaction and weakens superconductivity, which could be tested experimentally by doping a superconductor. Our work reveals the ambivalent relationship between non-interacting quantum geometry and superconductivity, and possibly in other correlated phases.  
\end{abstract}

\maketitle
\section*{Introduction}
In crystalline systems without electronic correlations the Bloch theorem states that the eigenstates are expressed as $\ket{\psi_n(\bm{k})}=e^{i\bm{k}\cdot\hat{\bm{r}}}\ket{u_n(\bm{k})}$, with $\ket{\psi_n(\bm{k})}$ the Bloch state and $\ket{u_n(\bm{k})}$ its cell-periodic counterpart \cite{Cayssol2021}. While the exponential factor is reminiscent of a free electron, the cell-periodic Bloch state is a consequence of the crystalline potential, i.e. the Coulomb interaction between an electron and the atomic nuclei composing the lattice. The Bloch states can be seen as representing quasiparticles, hereafter called Bloch fermions, which emerge from the electron-nuclei Coulomb interaction, analogously to quasiparticles in Fermi liquid theory which emerge from the electron-electron Coulomb interaction. For example, certain materials harbor Bloch states that mimic Dirac and Weyl fermions at low-energy \cite{Bradlyn2016}. The cell-periodic Bloch state encapsulates the emergent character of the Bloch fermion.  Bloch's theorem then establishes that the system of independent electrons in a crystalline lattice is formally equivalent to a gas of free quasiparticles, whose quasi-character is a remnant of the electron-nuclei Coulomb interaction. With the advent of topological and geometrical extensions of band theory, it is now well-established that the cell-periodic Bloch state can have significant influence on the properties of solids \cite{Qi2011,Cayssol2021,Törma2023,Liu2024,Yu2025}.  

The distinction between elementary electrons and Bloch fermions is especially relevant when discussing electronically correlated systems. As Bloch fermions differ from elementary electrons, interacting Bloch fermions also a priori differ from interacting electrons. Approximating Bloch fermions as electrons in an interacting problem neglects the effects of quantum geometry of the normal (non-interacting) state (NS) on the correlated phase. However, it is now known that the NS quantum geometry can significantly influence correlated phases. For example, the NS geometry is essential to fractional Chern insulators, zero-magnetic field analogues of fractional quantum Hall phases, where the geometry's uniformity in the Brillouin zone is instrumental for its stabilization \cite{Roy2014,Claassen2015}. Effects of the NS quantum geometry have also been found in excitons \cite{Srivastava2015,Zhou2015,Hichri2019,Cao2021} and superconductors \cite{Peotta2015,Rossi2021,Trm2022,Simon2022,Kitamura2022,Iskin2023,Porlles2023,Kitamura2023,Chen2024,Kitamura2024,Daido2024,Yu2024,Wang2024,Hu2025b,Li2025,Thumin2025}. In the present work, we focus on two such effects. On the one hand it was discovered that the NS quantum metric is a source of supercurrent, hereafter referred to as the geometric supercurrent, that is insensitive to whether the associated band is flat or dispersive \cite{Peotta2015,Rossi2021}. Quantum geometry is known to be related to the spread of Wannier functions through the localisation tensor, whose gauge-invariant part is the integral of the quantum metric over the Brillouin zone \cite{Marzari1997,Resta2011}. This has been used to interpret the geometric supercurrent as coming from an interaction-driven disruption of the Wannier functions' destructive interferences, thereby allowing the interacting particles to move \cite{Trm2022,Liu2024}. On the other hand, it was argued that the NS Berry curvature lowers the attractive interaction through an emergent Darwin term, thereby weakening the superconducting phase \cite{Simon2022}. These works point towards an ambiguous relationship between the NS quantum geometry and superconductivity, through a metric-curvature competition.  
    
In this work, we explore this ambivalent relation of the NS quantum geometry towards superconductivity based on the non-locality and emergence of Bloch fermions. First, we develop a momentum-resolved relation between quantum geometry and the quantum fluctuations of the position operator. The quantum geometry then appears as a quantifier of the non-locality of the Bloch fermion, itself a consequence of virtual interband transitions as well as the emergence and quasiparticle character of Bloch fermions. Starting from the idea that the Cooper pairing happens between Bloch fermions instead of elementary electrons \cite{Daido2024}, we then interpret the two effects mentioned above. First, the two contributions to the superfluid weight are associated with two independent movements of the Bloch fermions. The conventional supercurrent is associated to their center-of-mass motion, which disappears in flat band systems, while the geometric supercurrent is a result of their zero-point motions, which persists in flat-band systems. Second, we show that the same non-locality generates an emergent Darwin term which screens the pairing interaction, and thus weakens superconductivity. Therefore while the non-locality of the NS charge carriers unlocks another type of supercurrent, it can also be detrimental to interaction effects. The two effects we discussed thus appear as two sides of the same coin, with non-locality as the generator of this ambivalent relationship between the NS quantum geometry and superconductivity. 

The paper is organised as follows. In Section \ref{Sec:Qgeometry}, we briefly introduce multiparameter quantum estimation theory and discuss its application to band theory. In Section \ref{Sec:Sup-weight}, the conventional and geometric sources of supercurrent are interpreted based on Section \ref{Sec:Qgeometry}. In Section \ref{Sec:Darwin-term}, we phenomenologically derive a generalized form of the emergent Darwin term in one and two-body problems. We then compute the effective pairing potential in $s$-wave superconductors, and discuss the example of two-dimensional massive Dirac fermions.

\section{Emergence, non-locality and quantum geometry}
\label{Sec:Qgeometry}
In this section we first briefly review some aspects of multiparameter quantum estimation theory, and apply it to band theory. 
    \subsection{Multiparameter quantum estimation theory}
    \label{Ssec:Discussion-metro}

Multiparameter quantum estimation theory aims at finding optimal experimental protocols as well as tight error bounds, when multiple parameters are simultaneously estimated. References to the field include Refs. \cite{Boixo2007,Pang2014,Guo2016,Ragy2016,Liu2019,Carollo2019,Xing2020,Lu2021,Xia2023,Lambert2023}. For our discussion, it is sufficient to consider pure states. Let $\ket{\psi_0}$ be a quantum state representing a probe system, then parametrised using a unitary matrix $U(\bm{\theta})$ into $\ket{\psi(\bm{\theta})}=U(\bm{\theta})\ket{\psi_0}$ \cite{Guo2016,Boixo2007,Liu2019}. If the probe state is an eigenstate of the Hamiltonian $H_0$, then the parametrised state is an eigenvector of $H(\bm{\theta})=U(\bm{\theta})HU(\bm{\theta})^{\dagger}$.

 The optimal error achievable in the estimation of the parameters $\bm{\theta}$ is given by the inverse of the quantum Fisher information matrix (QFIM) $\mathcal{F}_{\mu\nu}(\bm{\theta})$, through the quantum Cramér-Rao bound \cite{Liu2019,Lambert2023} 
\begin{equation}
    \Cov(\hat{\theta}_\mu,\hat{\theta}_\nu)\geq\frac{1}{\mathcal{F}_{\mu\nu}}.
\end{equation}
The operators $\hat{\bm{\theta}}$ are estimators of the parameters $\bm{\theta}$, and we have made the $\bm{\theta}$-dependences implicit. For pure states, the QFIM reads \cite{Lambert2023,Liu2019}
\begin{subequations}
\begin{align}
    \mathcal{F}_{\mu\nu}&=4\Re[\bra{\partial_{\mu}\psi}\ket{\partial_{\nu}\psi}-\bra{\partial_{\mu}\psi}\ket{\psi}\bra{\psi}\ket{\partial_{\nu}\psi}]\\
    &=4g_{\mu\nu},
\end{align}
\end{subequations}
with $\partial_{\mu}=\partial_{\theta_\mu}$ and $g_{\mu\nu}=\Re Q_{\mu\nu}$ the quantum metric. Another formulation of the QFIM may be found by considering the generators of the parameters $\theta_{\mu}$, defined as $\hat{G}_{\mu}=i(\partial_{\theta_{\mu}}U^{\dagger})U$. It can be shown that the QFIM becomes
\begin{subequations}
\begin{align}
\mathcal{F}_{\mu\nu}&=4\Re\big[\langle \hat{G}_\mu \hat{G}_\nu\rangle_{\psi_0}-\langle \hat{G}_\mu\rangle_{\psi_0}\langle \hat{G}_\nu\rangle_{\psi_0}\big]\\
&=4\Re\Cov_{\ket{\psi_0}}(\hat{G}_\mu,\hat{G}_\nu),
\end{align}
\end{subequations}
with $\langle \hat{G}_{\mu}\rangle_{\psi_0}=\bra{\psi_0}\hat{G}_\mu\ket{\psi_0}$  \cite{Guo2016,Xia2023,Lambert2023}. The QFIM, and thus the quantum metric, is a consequence and quantifier of the quantum fluctuations of the parameter generators. In particular, the square root of the diagonal element $\mathcal{F}_{\mu\mu}$ is equal to the standard deviation of $\hat{G}_\mu$. The Berry curvature $\mathcal{B}_{\mu\nu}=-2\Im Q_{\mu\nu}$ can also be expressed in terms of the generators \cite{Guo2016,Xia2023},
\begin{equation}
    \mathcal{B}_{\mu\nu}=i\bra{\psi_0}[\hat{G}_\mu,\hat{G}_\nu]\ket{\psi_0}.
\end{equation}
The Berry curvature therefore stems from the non-commutativity of the parameter generators in the probe state. Applying the Heisenberg uncertainty relation then gives \cite{Guo2016}
\begin{equation}
    \mathcal{F}_{\mu\mu}\mathcal{F}_{\nu\nu}\geq 4\mathcal{B}_{\mu\nu}^2.
\end{equation}
The particularity of the multiparameter case is that the generators of the different parameters $\theta_{\mu}$ may not commute, which drives a non-zero Berry curvature that acts as a lower bound on the QFIM. In that case the associated parameters are said to be incompatible, since the non-trivial Berry curvature implies that they cannot be simultaneously and perfectly estimated \cite{Guo2016,Xia2023,Lu2021}.
    \subsection{Application to band theory}
The Bloch Hamiltonian is defined as
\begin{equation}
    H(\bm{k})=e^{-i\bm{k}\cdot\hat{\bm{r}}}He^{i\bm{k}\cdot\hat{\bm{r}}},
\end{equation}
with its eigenstates being the cell-periodic Bloch states $\ket{u_n(\bm{k})}=e^{-i\bm{k}\cdot\bm{\hat{r}}}\ket{\psi_n(\bm{k})}$ \cite{Cayssol2021}. Seeing this definition as a parametrisation process, the correspondence between band theory and multiparameter quantum estimation theory is then shown in Table \ref{Table:Correspondance}. The probe Hamiltonian is $H$, while its parametrized counterpart is the Bloch Hamiltonian $H(\bm{k})$. This implies that the estimated parameters are the components of the crystalline momentum $\bm{k}$, and that the parametrizing unitary matrix is $U(\bm{k})=e^{-i\bm{k}\cdot\hat{\bm{r}}}$. The probe state is then given by the Bloch state $\ket{\psi_n(\bm{k})}$, and the analogue of $\ket{\psi(\bm{\theta})}$ is the cell-periodic Bloch state $\ket{u_n(\bm{k})}$.
\begin{table}[h!]
\centering
\begin{tabular}{| c | c |}
\hline Estimation theory & Band theory\\\hline  $\ket{\psi_0}$ & $\ket{\psi_n(\bm{k})}$ \\\hline  $\bm{\theta}$ & $\bm{k}$\\\hline  $U(\bm{\theta})$ & $\exp\big(-i\bm{k}\cdot\hat{\bm{r}}\big)$\\\hline  $\ket{\psi(\bm{\theta})}$ & $\ket{u_n(\bm{k})}$\\\hline
\end{tabular}
\caption{Correspondence between band theory and multiparameter quantum estimation theory.}
\label{Table:Correspondance}
\end{table}
We show in appendix \ref{Appendix:QGT-pos} that the quantum geometric tensor (QGT) indeed takes the form of a covariance matrix,
\begin{equation}
    Q^{n}_{\mu\nu}=\Cov_{\ket{\psi_n}}(\hat{R}_{\mu},\hat{R}_{\nu}),
    \label{eq:QGT-Covpos}
\end{equation}
with the generator of $k_{\mu}$ defined as $\hat{R}_{\mu}=-i(\partial_{\mu}U^{\dagger})U$. Note that the relation between the QGT and the generators is mentioned in Ref. \cite{Provost1980}. As shown in appendix \ref{Appendix:QGT-pos}, the generator can be further expressed as follows,
\begin{equation}
    \hat{R}_{\mu}=\int_{0}^{1}e^{is\bm{k}\cdot\hat{\bm{r}}}\hat{r}_{\mu}e^{-is\bm{k}\cdot\hat{\bm{r}}}\dd s=\sum_{n=0}^{+\infty}\frac{i^n}{(n+1)!}\operatorname{ad}^{n}_{\bm{k}\cdot\hat{\bm{r}}}\hat{r}_{\mu},
    \label{eq:Generator-position}
\end{equation}
with $\ad_{A}$ the adjoint representation, which obeys $\operatorname{ad}_{A}B=[A,B]$ and $\operatorname{ad}^{n}_{A}B=[A,\operatorname{ad}^{n-1}_{A}B]$. Making use of the closure relation for the position states $\ket{\bm{r}}$, we get
\begin{subequations}
    \begin{align}
        \hat{R}_{\mu}&=\int_{0}^{1}\dd s\int\dd\bm{r}e^{is\bm{k}\cdot\bm{r}}r_{\mu}e^{-is\bm{k}\cdot\bm{r}}\ket{\bm{r}}\bra{\bm{r}}\\
        &=\int\dd\bm{r}r_{\mu}\ket{\bm{r}}\bra{\bm{r}}=\hat{r}_{\mu}.
    \end{align}
\end{subequations}
The QGT then becomes the covariance matrix of the position operator in the Bloch state.  Since the Bloch fermion is represented by the Bloch state, quantum geometry is then directly related to the quantum fluctuations of the position of the Bloch fermion, i.e. its non-locality. When pictured over time, this non-locality may be viewed as the zero-point motions of the Bloch fermion.  In particular, the diagonal elements of the quantum metric and the Berry curvature read
\begin{align}
    g^{n}_{\mu\mu}(\bm{k})&=\bra{\psi_n(\bm{k})}\hat{r}_{\mu}^2\ket{\psi_n(\bm{k})}-\bra{\psi_n(\bm{k})}\hat{r}_{\mu}\ket{\psi_n(\bm{k})}^2,\nonumber\\
    \mathcal{B}^{n}_{\mu\nu}(\bm{k})&=i\bra{\psi_n(\bm{k})}[\hat{r}_{\mu},\hat{r}_{\nu}]\ket{\psi_n(\bm{k})}.
    \label{eq:Qgeom-pos}
\end{align}
 Quantum geometry has been previously linked with non-locality. On the one hand, the integral of the quantum metric over the Brillouin zone gives the invariant part of the quadratic spread of the Wannier functions \cite{Marzari1997,Resta2011}. This fact has been used to formulate a time-dependent extension of the integrated QGT \cite{Komissarov2024,Yu2025}. On the other hand, the (intraband) position operator can be represented in the basis of Bloch states by $\hat{r}^{n}_{c\mu}=i\partial_{\mu}+\mathcal{A}_{n\mu}$, with $\bm{\mathcal{A}}_n$ the Berry connection \cite{Blount1962}. One then finds $[\hat{r}^{n}_{c\mu},\hat{r}^{n}_{c\nu}]=i\mathcal{B}^{n}_{\mu\nu}$, analogously to Eq. (\ref{eq:Qgeom-pos}). In contrast to these two approaches, Eq. (\ref{eq:Qgeom-pos}) is a momentum-resolved link between non-locality and quantum geometry that directly involves the position operator of the Bloch states.
\begin{figure}[t!]
    \centering
    \includegraphics[width=0.85\linewidth]{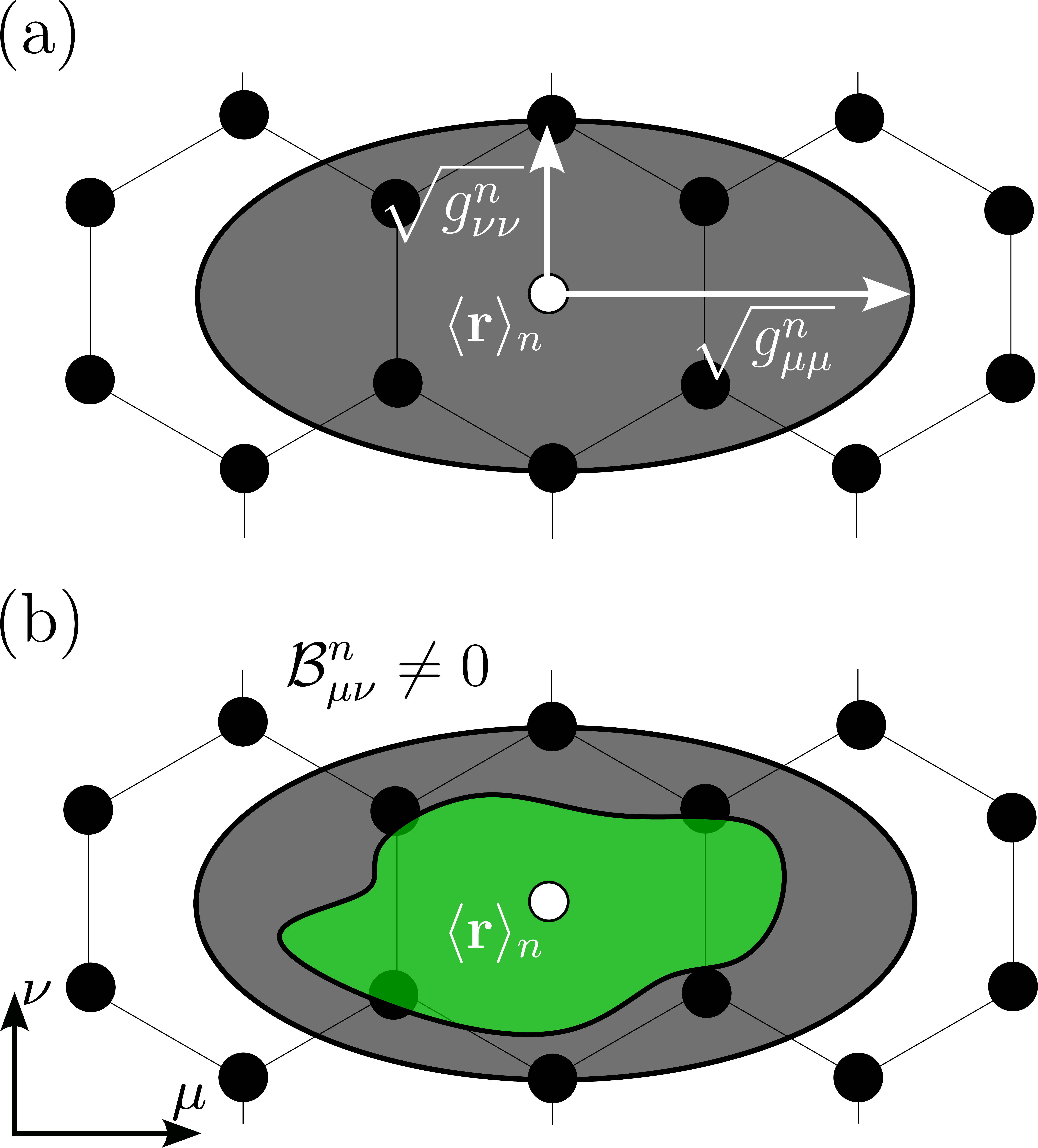}
    \caption{Relation between the non-locality of Bloch fermions and quantum geometry in the $(\mu,\nu)$ plane, visualized on the honeycomb lattice. (a) The product of the square root of the quantum metric elements defines a typical area of non-locality around the average position, shown in grey. (b) A non-trivial Berry curvature defines a minimal area of non-locality of the Bloch fermion, shown in green.}
    \label{fig:Non-local-wavepacket}
\end{figure}

From Eq. (\ref{eq:Qgeom-pos}), $g^{n}_{\mu\mu}$ is the standard deviation of the position in the $\mu$-direction, and as such defines a length $\sqrt{g^{n}_{\mu\mu}}$ which quantifies the Bloch fermion's \textit{typical} non-locality around the average position $\langle\bm{r}\rangle_{n}=\bra{\psi_n(\bm{k})}\hat{\bm{r}}\ket{\psi_n(\bm{k})}$. In the two-dimensional plane $(\mu,\nu)$  the Bloch fermion has a typical area of non-locality given by $\mathcal{A}^{n}_{\text{typ},\mu\nu}=\pi\sqrt{g^{n}_{\mu\mu}g^{n}_{\nu\nu}}$, as pictured in Fig.~\ref{fig:Non-local-wavepacket}a. Contrastingly, the Berry curvature appears as the commutator of different elements of the position operator in the Bloch state. The Berry curvature is non-trivial only if the position operators do not mutually commute in the Bloch state. As the elementary electrons obey the mutual commutativity of the position operators, this highlights the emergent character and quasiparticle nature of the Bloch fermion, and its importance for the existence of band topology.
Applying the Heisenberg uncertainty relation, as done in Sec. \ref{Ssec:Discussion-metro}, gives
\begin{equation}
    \sqrt{g^{n}_{\mu\mu}(\bm{k})g^{n}_{\nu\nu}(\bm{k})}\geq\frac{1}{2}|\mathcal{B}^{n}_{\mu\nu}(\bm{k})|.
    \label{eq:Uncertainty-rel-band}
\end{equation}

 The Berry curvature then puts a lower bound on the typical area of non-locality, thus defining a \textit{minimal} area of non-locality $\mathcal{A}^{n}_{\text{min},\mu\nu}=\frac{\pi}{2}|\mathcal{B}^{n}_{\mu\nu}|$ as shown in Fig.~\ref{fig:Non-local-wavepacket}b. Eq.~(\ref{eq:Uncertainty-rel-band}) can be independently derived by using the semi-positive definiteness of the QGT \cite{Roy2014,Yu2025}. 

The emergent non-locality of the Bloch fermion is reminiscent to that of the electron in relativistic quantum mechanics. Such electrons undergo erratic virtual electron-positron transitions with a typical length given by the Compton wavelength $\lambdabar_c=\hbar/mc$, where $m$ is the electron's mass and $c$ the speed of light. These virtual processes then outline a region of space in which the electron is not localised, and relativistic effects must be taken into account \cite{Zelevinsky2010}. In the case of Bloch states, virtual interband processes, known to be the source of quantum geometry \cite{Cayssol2021}, outline an region of space in which the Bloch state cannot be localised. This analogy is used in Sec. \ref{Sec:Darwin-term} to derive a generalized form of the Darwin term previously found in the case where Bloch fermions are two-dimensional massive Dirac fermions \cite{Simon2022}.

\section{Superfluid weight interpretation}
\label{Sec:Sup-weight}
Supercurrents formally originate from the electrodynamic response of the superconductor through the London equation, which in the static limit is encapsulated in the superfluid weight \cite{Peotta2015,Liang2017,Trm2022}. Beyond a previously known contribution, dubbed conventional, coming from the momentum derivatives of the band dispersion \cite{Chandrasekhar1993}, it was shown that the superfluid weight contains a contribution driven by the derivatives of the cell-periodic Bloch states \cite{Peotta2015}. In particular, when the band of interest is distanced enough in energy from others, the so-called isolated band limit, this geometric contribution is shown to be driven by the normal state quantum metric. Since this contribution is independent of the momentum derivatives of the band dispersion, it persists in flat-band systems and has since garnered significant interest \cite{Liang2017,Rossi2021,Trm2022,Tian2023,Kitamura2022b,Daido2024}. 

In this section, we discuss the interpretation of the conventional and geometric superfluid weight, and the associated supercurrents, based on Sec. \ref{Sec:Qgeometry}. The key point in our argument is that the Cooper pairs are not formed by electrons, but by Bloch fermions that have the emergent property of having a non-locality quantified by the NS quantum geometry. As transport stems from a movement of charge, the two contributions to the superfluid weight, and the associated supercurrents, are then associated to two independent movements of the Bloch fermions forming the Cooper pairs.

\begin{figure}[t!]
    \centering
    \includegraphics[width=0.85\linewidth]{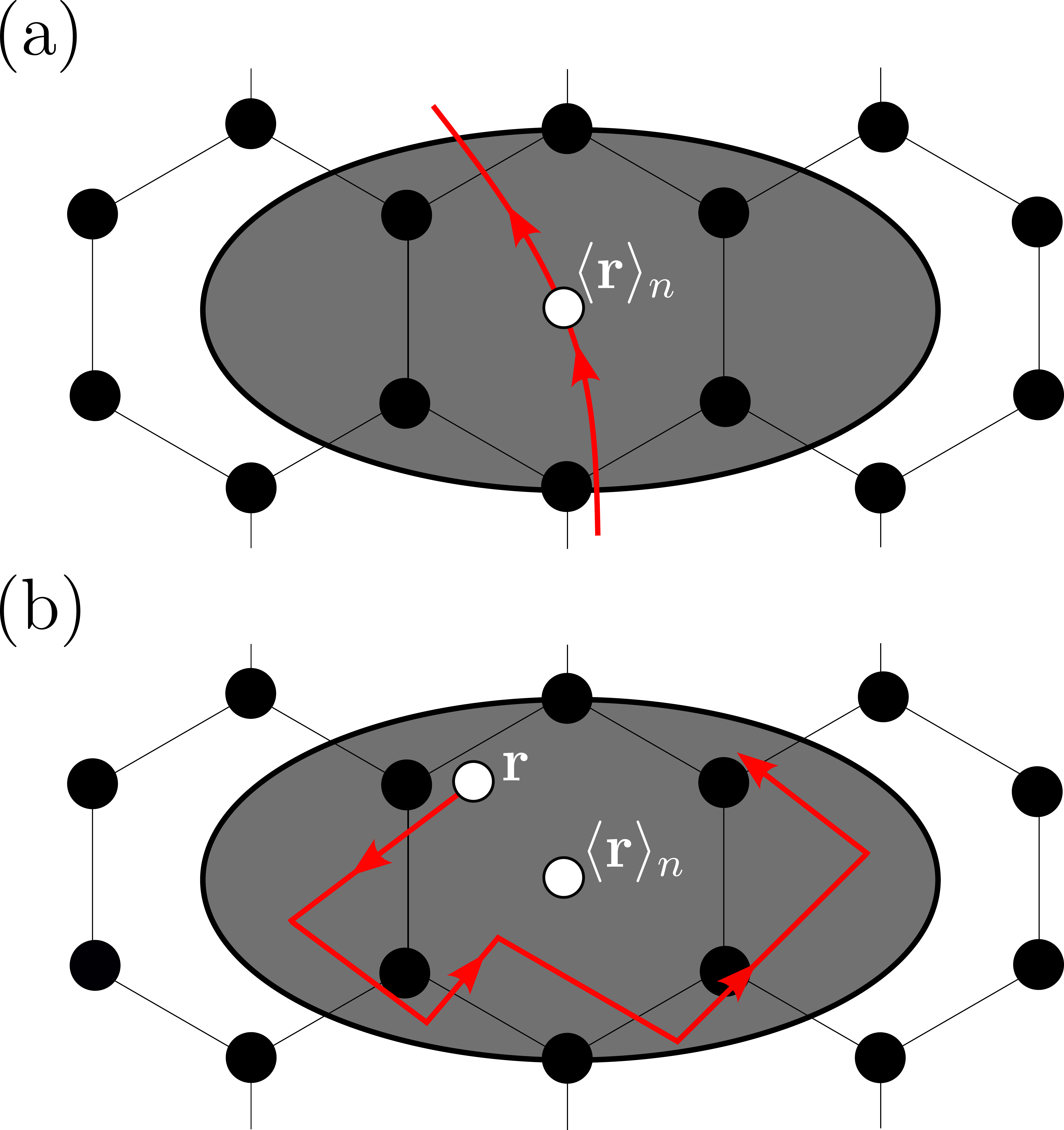}
    \caption{(a) Overall movement of the Bloch fermion, associated to the conventional superfluid weight (b) Zero-point motion of the Bloch fermion, associated to the geometric superfluid weight.
    }
    \label{fig:Movements-Super-weight}
\end{figure}
    \subsection{Conventional supercurrent}
    For a superconducting gap $\Delta$, the conventional contribution to the superfluid weight of the $n$-th band reads \cite{Liang2017}
    \begin{equation}
        D^{n}_{\text{conv},\mu\nu}=\sum_{\bm{k}}\frac{\Delta^2}{E_{n}^3}\partial_{\mu}\epsilon_n\partial_{\nu}\epsilon_n,
    \end{equation}
    where $E_n=\sqrt{(\epsilon_n-\mu)^2+\Delta^2}$ with $\mu$ the chemical potential, and for simplicity we consider the zero temperature limit. The conventional superfluid weight, and the subsequent conventional supercurrent, is thus driven by the momentum derivatives of the band dispersion. Since the time derivative of the average position $\langle\hat{\bm{r}}\rangle_n$ is given by $\partial_{\bm{k}}\epsilon_{n}$, the conventional supercurrent can then be interpreted as a result of center-of-mass motion of the Bloch fermions, as pictured in Fig.~\ref{fig:Movements-Super-weight}a.

    In the case of flat band systems where $\epsilon_n(\bm{k})$ is a constant, the average position of the Bloch fermion stays still. If the wavepacket is taken to be local, then the charge carriers of flat band systems are immobile. However as a non-trivial quantum geometry implies that they are non-local, they only stay immobile \textit{as a whole}.
    \subsection{Geometric supercurrent}
    In the isolated band limit, the geometric contribution to the superfluid weight of the $n$-th band is given by
    \begin{equation}
        D^{n}_{\text{geom},\mu\nu}=2\Delta^2\sum_{\bm{k}}\frac{1}{E_{n}}g^{n}_{\mu\nu},
    \end{equation}
    where once again for simplicity we consider the zero temperature limit, and $\Delta$ is the superconducting gap. In Sec. \ref{Sec:Qgeometry} we saw that the quantum metric quantifies the typical non-locality of Bloch fermions, which can be seen as its zero-point motion when pictured over time. As such, we interpret the geometric contribution to the superfluid weight as stemming from the zero-point motions of the Bloch fermions forming the Cooper pairs. For flat band systems, although the Bloch fermions stay still as a whole, their true position still jitters around the average position through quantum fluctuations, as pictured in Fig.~\ref{fig:Movements-Super-weight}b. The geometric supercurrent thus also appears as a quantum mechanical and uncertainty-driven supercurrent.

    Our interpretation contrasts with a previous argument, which can be summarized as follows \cite{Trm2022,Yu2025}. A sizeable quantum metric yields substantially spread Wannier functions, which interfere destructively without interactions, such that the single-particle states are immobile. The addition of the pairing interaction then disrupts these interferences such that the interacting particles, the Cooper pairs, can move. The geometric supercurrent is then seen as an interaction-driven mechanism. However, we have argued that the geometric supercurrent stems from the zero-point motion of the non-interacting (in terms of electron-electron interaction) particles forming the Cooper pairs. Therefore, the geometric supercurrent solely stems from a property of the normal state that is inherited by the superconducting state. Our interpretation is in accordance with the recent discovery that the geometric superfluid weight stems from the band-projected pairing matrix, and is thus a normal state property influencing superconductivity \cite{Daido2024}. The geometric contribution to the superfluid weight was also recently interpreted in terms of virtual interband transitions \cite{Hu2025a}, which is consistent with our results since virtual interband transitions produce non-locality and quantum geometry \cite{Cayssol2021}.

    Lastly, we focus on the topological and geometric bound on the geometric superfluid weight \cite{Peotta2015,Trm2022}. Indeed, using the results of Sec. \ref{Sec:Qgeometry} we can rederive the geometric and topological bounds on the geometric superfluid weight and formulate a physical interpretation. Young's inequality together with Eq.~(\ref{eq:Uncertainty-rel-band}) imply that
    \begin{equation}
        \frac{1}{2}\big(g^{n}_{\mu\mu}+g^{n}_{\nu\nu}\big)\geq\sqrt{g^{n}_{\mu\mu}g^{n}_{\nu\nu}}\geq\frac{1}{2}|\mathcal{B}^{n}_{\mu\nu}|,
    \end{equation}
    from which we have $g^{n}_{\mu\mu}+g^{n}_{\nu\nu}\geq|\mathcal{B}^{n}_{\mu\nu}|$. For two-dimensional systems, this reduces to the inequality $\Tr g^{n}\geq|\mathcal{B}^{n}_{xy}|$. We thus find the geometric bound
    \begin{equation}
        D^{n}_{\text{geom},\mu\mu}+D^{n}_{\text{geom},\nu\nu}\geq\sum_{\bm{k}}\frac{2\Delta^2}{E_n}|\mathcal{B}^{n}_{\mu\nu}|.
    \end{equation}
    For two-dimensional isotropic systems where $D^{n}_{\mu\nu}=D^{n}\delta_{\mu\nu}$, one finds the geometric bound derived in Ref. \cite{Liang2017}. In particular, if the $n$-th band is exactly flat then using the triangular inequality, and moving the summation to an integration, we find
    \begin{equation}
        D^{n}_{\text{geom}}\geq\frac{2\Delta^2}{E_n}\bigg|\int_{\text{BZ}}\mathcal{B}^{n}_{xy}\frac{\dd^2\bm{k}}{(2\pi)^2}\bigg|=\frac{4\pi\Delta^2}{E_n}|\mathcal{C}_n|,
    \end{equation}
    which is the known topological bound on the geometric superfluid weight \cite{Peotta2015,Liang2017}.

    The fact that the geometric superfluid weight is bounded from below by the Berry curvature can be interpreted as follows. As argued in Sec. \ref{Sec:Qgeometry}, Eq.~(\ref{eq:Uncertainty-rel-band}) implies that the Berry curvature sets a minimal area of non-locality of the Bloch fermion. It therefore also sets a minimal amount of zero-point motion. As such, the Berry curvature then ensures (or protects) a minimal amount of geometric supercurrent.

\section{Emergent Darwin term}
\label{Sec:Darwin-term}
 The Darwin term first surfaced as an object of study in superconductivity following efforts to formulate a relativistic theory of superconductivity \cite{Bailin1982,Capelle1995,Strange1998,Capelle2001}. Analogously, an \textit{emergent} Darwin term also arises from the quantum geometry carried by Bloch fermions \cite{Gosselin2010,Zhou2015,Srivastava2015,Simon2022}. It has been derived via an $\hbar$-expansion of a generic Hamiltonian \cite{Gosselin2010}, and was later used to explain non-hydrogenic spectra of excitons \cite{Zhou2015,Srivastava2015}. One notable difference is that the Darwin term in Ref. \cite{Gosselin2010} is linked to the quantum metric whereas the Darwin term in Refs. \cite{Zhou2015,Srivastava2015} is based on the Berry curvature. In line with Refs. \cite{Zhou2015,Srivastava2015}, Ref. \cite{Simon2022} derived a Berry curvature-based Darwin term to study Berry curvature effects in superconductivity.
 
In this section we generalise the emergent Darwin term previously found in Ref. \cite{Simon2022}, with the non-locality of the Bloch fermion as its driver. Our derivation is based on the phenomenological derivation presented in Ref. \cite{Zelevinsky2010} for the elementary electron in relativistic quantum mechanics. We thus find a Darwin term driven by the quantum metric, as in Ref. \cite{Gosselin2010}. Moreover, we do so through a physically transparent phenomenological derivation which is then naturally extended to a generic two-body problem. We also show that the quantum metric-based Darwin term reduces to the Berry curvature-based one in the case of low-energy massive Dirac fermions close to the band gap, considered in Refs. \cite{Zhou2015,Srivastava2015,Simon2022}.
    \subsection{One-body problem}
    \label{Ssec:Darwin-1corps}
Consider a Bloch state $\ket{\psi_n(\bm{k})}$ with the associated band dispersion $\epsilon_n(\bm{k})$, subject to an external potential $V(\hat{\bm{r}})$. The Hamiltonian of interest is given by
\begin{equation}
    H=\epsilon_n\ket{\psi_n}\bra{\psi_n}+V(\hat{\bm{r}}),
\end{equation}
where the $\bm{k}$-dependences are implicit. In the following, for an operator $\hat{\mathcal{O}}$ we denote $\langle\hat{\mathcal{O}}\rangle_{n}=\bra{\psi_n}\hat{\mathcal{O}}\ket{\psi_n}$. The projected Hamiltonian  then reads
\begin{equation}
    \langle H\rangle_n=\epsilon_n+\bra{\psi_n}V(\hat{\bm{r}})\ket{\psi_n}.
\end{equation}
As the Bloch fermion is non-local, the position operator fluctuates around its average $\langle\hat{\bm{r}}\rangle_n$. To second order in the position fluctuation $\delta\hat{r}_\mu=\hat{r}_{\mu}-\langle\hat{r}_{\mu}\rangle_n$, we obtain
\begin{equation}
    \langle H\rangle_n=\epsilon_n+V(\langle\hat{\bm{r}}\rangle_n)+\frac{1}{2}\langle\delta \hat{r}_{\mu}\delta \hat{r}_{\nu}\rangle_n\partial_{\mu}\partial_{\nu}V(\langle\hat{\bm{r}}\rangle_n),
\end{equation}
adopting the Einstein summation convention, and where $\partial_{\mu}=\partial_{r_{\mu}}$. The first order term vanishes by linearity of the average $\langle\cdot\rangle_n$. From Sec.~\ref{Sec:Qgeometry}, we have $\langle\delta\hat{r}_{\mu}\delta\hat{r}_{\nu}\rangle_n=Q^{n}_{\mu\nu}$. Additionally, assuming that the potential $V$ is twice continuously differentiable around $\langle\hat{\bm{r}}\rangle_n$, the resulting symmetry of partial derivatives under permutation yields
\begin{equation}
    \langle H\rangle_n=\epsilon_n+V(\langle\hat{\bm{r}}\rangle_n)+\frac{1}{2}g^{n}_{\mu\nu}\partial_{\mu}\partial_{\nu}V(\langle\hat{\bm{r}}\rangle_n).
\end{equation}
Going to the local approximation $\langle\hat{\bm{r}}\rangle_n\mapsto\bm{r}$ then finally yields the effective Hamiltonian containing the emergent Darwin term,
\begin{equation}
    H_{\text{eff},n}=\epsilon_n+V(\bm{r})+\frac{1}{2}g^n_{\mu\nu}\partial_{\mu}\partial_{\nu}V(\bm{r})
    \label{eq:1body-Darwinterm}
\end{equation}

As mentioned in Sec. \ref{Sec:Qgeometry}, the physical origin of the emergent Darwin term is analogous to that of elementary electrons in relativistic quantum mechanics \cite{Zelevinsky2010}. Bloch fermions undergo virtual interband transitions over a typical length quantified by quantum geometry, which blurs their position over an associated area. Specifically, the emergent Darwin term is driven by the Bloch fermion's typical non-locality, and thus by the quantum metric. Contrastingly, the previously found expression of the Darwin term in terms of the Berry curvature is associated to the minimal non-locality of the Bloch fermion \cite{Simon2022}. 

For two-dimensional massive Dirac fermions at the $K$ point, whose Bloch Hamiltonian and quantum geometry are summarised is appendix \ref{annexe:Massive-Dirac}, Eq.~(\ref{eq:1body-Darwinterm}) becomes
\begin{equation}
    H_{\text{eff},n}=\epsilon_n+V(\bm{r})+\frac{\lambdabar_c^2}{4}\Delta V(\bm{r}),
\end{equation}
with $\lambdabar_c$ the reduced Compton wavelength, explicited in appendix \ref{annexe:Massive-Dirac}, and $\Delta$ the Laplacian operator. The Darwin term in this case differs by a factor of two from the full expression obtained by means of a Foldy-Wouthuysen transformation \cite{Zelevinsky2010,Simon2022}. This discrepancy is also present in the case of elementary electrons in relativistic quantum mechanics, and originates from the phenomenological character of the derivation \cite{Zelevinsky2010}.

Omitting this discrepancy, the peculiarity of two-dimensional massive Dirac fermions is that its quantum geometry is ideal at the $K$ point \cite{Roy2014}, as shown in appendix \ref{annexe:Massive-Dirac}. Indeed, at the $K$ point we have $\Tr g^{n}=|\mathcal{B}^n_{xy}|$ and $\sqrt{\det g^{n}}=\sqrt{g^{n}_{xx}g^{n}_{yy}}=|\mathcal{B}^{n}_{xy}|/2$. In that sense, the ideality of the quantum geometry implies that for two-dimensional massive Dirac fermions, the typical non-locality coincides with the minimal one, and as such is optimal. 
    
    \subsection{Two-body problem}
We note that the relation between the NS quantum geometry and superconductivity in the two-body problem has previously been considered for the superfluid weight \cite{Törmä2018,Iskin2022,Iskin2024}, and the Cooper pair size \cite{Iskin2025}. We focus here on the role of the emergent Darwin term. For the two-body problem, we consider two Bloch states $\ket{\psi_n(\bm{k}_1)}$ and $\ket{\psi_m(\bm{k}_2)}$ with the associated band dispersions $\epsilon_{n}(\bm{k}_1)$ and $\epsilon_{m}(\bm{k}_2)$, respectively. The respective Hilbert spaces are $\mathscr{H}_{1}$ and $\mathscr{H}_2$, such that the full Hilbert space is $\mathscr{H}=\mathscr{H}_1\otimes\mathscr{H}_2$. In the following, for convenience we denote operators acting on $\mathscr{H}$ of the form $\mathcal{O}\otimes\mathbbm{1}$ and $\mathbbm{1}\otimes\mathcal{O}$ as $\mathcal{O}_1$ and $\mathcal{O}_2$, respectively. Let $\mathcal{P}_{n}(\bm{k})=\ket{\psi_n(\bm{k})}\bra{\psi_n(\bm{k})}$ be the projector on the Bloch state $\ket{\psi_n(\bm{k})}$. The Bloch states are then subject to a translation-invariant potential V, depending only on $\hat{\bm{\rho}}=\hat{\bm{r}}_1-\hat{\bm{r}}_2$. The two-body Hamiltonian is then defined as
\begin{equation}
    H_2=\epsilon_{n}\mathcal{P}_{n1}+\epsilon_{m}\mathcal{P}_{m2}+V(\hat{\bm{\rho}}),
\end{equation}
where the $\bm{k}$-dependences are implicit. Define the two-body state $\ket{\Psi}=\ket{\psi_n(\bm{k}_1)}\otimes\ket{\psi_m(\bm{k}_2)}$. The projected Hamiltonian $\langle H_{2}\rangle_{\Psi}=\bra{\Psi}H_{2}\ket{\Psi}$ reads
\begin{equation}
    \langle H_{2}\rangle_{\Psi}=\epsilon_{n}+\epsilon_{m}+\bra{\Psi}V(\hat{\bm{\rho}})\ket{\Psi}.
\end{equation}
The average value of $\hat{\bm{\rho}}$ is given by
\begin{subequations}
    \begin{align}
        \langle\hat{\bm{\rho}}\rangle_{\Psi}&=\bra{\psi_n}\otimes\bra{\psi_m}\hat{\bm{r}}_1\otimes\mathbbm{1}-\mathbbm{1}\otimes\hat{\bm{r}}_2\ket{\psi_n}\otimes\ket{\psi_m}\\
        &=\langle\hat{\bm{r}}_1\rangle_n-\langle\hat{\bm{r}}_2\rangle_m,
    \end{align}
\end{subequations}
with $\langle\hat{\bm{r}}_2\rangle_{n}=\bra{\psi_n}\hat{\bm{r}}_1\ket{\psi_n}$ and $\langle\hat{\bm{r}}_2\rangle_{m}=\bra{\psi_m}\hat{\bm{r}}_2\ket{\psi_m}$. As in the one-body problem, we expand the potential $V$ to second order in the fluctuations $\delta\hat{\rho}_{\mu}=\hat{\rho}_{\mu}-\langle\hat{\rho}_{\mu}\rangle_{\Psi}$ and find
\begin{equation}
    \langle H_{2}\rangle_{\Psi}=\epsilon_{n}+\epsilon_{m}+V(\langle\hat{\bm{\rho}}\rangle_{\Psi})+\frac{1}{2}\langle\delta\hat{\rho}_{\mu}\delta\hat{\rho}_{\nu}\rangle_{\Psi}\partial_{\mu}\partial_{\nu}V(\langle\hat{\bm{\rho}}\rangle_{\Psi}),
\end{equation}
with $\partial_{\mu}=\partial_{\rho_{\mu}}$. It is shown in appendix \ref{annexe:2bodypb} that 
\begin{equation}
    \langle\delta\hat{\rho}_{\mu}\delta\hat{\rho}_{\nu}\rangle_{\Psi}=Q^{n}_{\mu\nu}+Q^{m}_{\mu\nu}.
    \label{eq:Fluct-2b-QG}
\end{equation}
We again assume that the potential $V$ is twice continuously differentiable around $\langle\hat{\bm{\rho}}\rangle_{\Psi}$, so that we have $(Q^{n}_{\mu\nu}+Q^{m}_{\mu\nu})\partial_{\mu}\partial_{\nu}=(g^{n}_{\mu\nu}+g^{m}_{\mu\nu})\partial_{\mu}\partial_{\nu}$. Going to the local approximation $\langle\hat{\bm{\rho}}\rangle_{\Psi}\mapsto\bm{\rho}$, the effective two-body Hamiltonian is given by
\begin{align}
    H_{\text{eff},2}=\epsilon_{n}(\bm{k}_1)&+\epsilon_{m}(\bm{k}_2)+V(\bm{\rho})\nonumber\\
    &+\frac{1}{2}(g^{n}_{\mu\nu}(\bm{k}_1)+g^{m}_{\mu\nu}(\bm{k}_2))\partial_{\mu}\partial_{\nu}V(\bm{\rho}),
\end{align}
where we have reintroduced the momentum dependence. The Darwin term of the two-body problem is thus simply given by the sum of the two one-body Darwin terms. We emphasize that the results obtained here are for a general potential, and therefore a priori concern any type of electronic correlation. We may also point out an important difference between the Darwin term considered for relativistic elementary electrons and its emergent counterpart for Bloch fermions. In the case of elementary electrons, $V$ is taken to be the Coulomb potential such that, in three dimensions, its Laplacian is $\delta(\bm{r})$ and only the $s$-wave state is affected. However, here $V$ may be the pairing potential, whose Laplacian is not given by the Dirac delta function. Therefore other types of pairing beyond $s$-wave may be affected, as shown in Ref. \cite{Simon2022}.

Focusing on the case  of superconductivity, where $V$ is a pairing potential, we assume spin is a good quantum number so that Bloch states are indexed as $\ket{\psi_{n\sigma}(\bm{k})}$. As a paradigmatic example we further consider pairing between time-reversed partners, i.e. of opposite spin and momenta, in the same band. Time-reversal symmetry implies that $g^{n\downarrow}_{\mu\nu}(-\bm{k})=g^{n\uparrow}_{\mu\nu}(\bm{k})\equiv g^{n}_{\mu\nu}(\bm{k})$ and $\epsilon_{n\downarrow}(-\bm{k})=\epsilon_{n\uparrow}(\bm{k})\equiv\epsilon_n(\bm{k})$, so that the effective two-body Hamiltonian becomes
\begin{equation}
    H_{\text{eff},2}=2\epsilon_{n}(\bm{k})+V(\bm{\rho})+g^{n}_{\mu\nu}(\bm{k})\partial_{\mu}\partial_{\nu}V(\bm{\rho})
    \label{eq:Darwin-Supra}
\end{equation}

\subsubsection{Cooper problem}
 Based on Eq.~(\ref{eq:Darwin-Supra}) we consider the Cooper problem of Bloch fermions within the same conduction band, and related by time-reversal, following Refs. \cite{Tinkham2004,Simon2022}. Furthermore, for simplicity, we consider $s$-wave pairing. Let $\Psi(\bm{\rho})$ be the pair wavefunction, with energy $E$. The pair wavefunction and the pairing potential have the following Fourier decomposition,
 \begin{equation}
     \Psi(\bm{\rho})=\sum_{\bm{k}}p_{\bm{k}}e^{i\bm{k}\cdot\bm{\rho}},\hspace{2mm}V(\bm{\rho})=\sum_{\bm{k}}V_{\bm{k}\bm{k'}}e^{i(\bm{k}-\bm{k'})\cdot\bm{\rho}}.
 \end{equation}
The eigenvalue problem with the effective Hamiltonian defined in Eq.~(\ref{eq:Darwin-Supra}) then yields \cite{Tinkham2004,Simon2022}
\begin{equation}
    (E-2\epsilon_{n}(\bm{k}))p_{\bm{k}}=\sum_{\bm{k'}}V^{\text{eff}}_{\bm{k}\bm{k'}}p_{\bm{k'}},
\end{equation}
where the effective interaction is given by
\begin{equation}
    V^{\text{eff}}_{\bm{k}\bm{k'}}=\Big[1-g^{n}_{\mu\nu}(\bm{k})(k_{\mu}-k'_{\mu})(k_{\nu}-k'_{\nu})\Big]V_{\bm{k}\bm{k'}}.
    \label{eq:Veff}
\end{equation}
The normal state quantum metric, and thus the non-locality of the Bloch fermions forming the Cooper pairs, seem to drive a screening of the pairing interaction through the emergent Darwin term. The extreme limit where the interaction is completely screened hints toward the existence of a \textit{geometric Pauli limit}. Note that the link between the QGT and the generators of the crystalline momentum has been invoked to interpret the effect of quantum geometry on energy shifts in excitonic spectra, but in terms of an emergent Lamb shift rather than a Darwin term \cite{Srivastava2015}.

Following the original Cooper problem, we find the following self-consistency equation \cite{Cooper1956,Tinkham2004}
\begin{equation}
    \sum_{\bm{k'}}\frac{\langle V^{\text{eff}}_{\bm{k}\bm{k'}}\rangle}{E-2\epsilon_{n}(\bm{k})}=1,
    \label{eq:self-cons}
\end{equation}
where the average $\langle\cdot\rangle$ is the weighted average with respect to the coefficients $p_{\bm{k}}$,
\begin{equation}
    \langle\mathcal{O}\rangle=\sum_{\bm{k'}}\mathcal{O}\frac{p_{\bm{k'}}}{\underset{\bm{k'}}{\sum}p_{\bm{k'}}}.
\end{equation}
We then expand the averaged effective interaction and get
\begin{widetext}
\begin{equation}
    \langle V^{\text{eff}}_{\bm{k}\bm{k'}}\rangle=\big[1-g^{n}_{\mu\nu}(\bm{k})k_{\mu}k_{\nu}\big]\langle V_{\bm{k}\bm{k'}}\rangle+2g^{n}_{\mu\nu}(\bm{k})k_{\mu}\langle k_{\nu}'V_{\bm{k}\bm{k'}}\rangle-g^{n}_{\mu\nu}(\bm{k})\langle k_{\mu}'k_{\nu}'V_{\bm{k}\bm{k'}}\rangle.
    \label{eq:Veff-expand}
\end{equation}
\end{widetext}

The pairing potential $V_{\bm{k}\bm{k'}}$ in the case of BCS $s$-wave superconductivity is  given by $V_{\bm{k}\bm{k'}}=-V\mathbbm{1}_{\mathcal{D}}(\bm{k})\mathbbm{1}_{\mathcal{D}}(\bm{k'})$ \cite{BCS57}. $\mathbbm{1}_{\mathcal{D}}$ is the indicator function of the set $\mathcal{D}$, which gathers all points of the Brillouin zone such that $\epsilon_F\leq\epsilon_n(\bm{k})\leq\epsilon_F+\hbar\omega_D$, with the Fermi level $\epsilon_F$ and the Debye energy $\hbar\omega_D$. Since the band dispersion for massive Dirac fermions is isotropic, the set $\mathcal{D}$ forms an annulus around the $k=0$ point. The summation/integration of the odd function $\bm{k}'\mapsto k_{\nu}'V_{\bm{k}\bm{k'}}$ will, by symmetry around $k'=0$, therefore vanish. The linear terms $\langle k_{\nu}'V_{\bm{k}\bm{k'}}\rangle$ are thus not relevant to the effective interaction, and using the expression of $V_{\bm{k}\bm{k'}}$ gives
\begin{equation}
    \langle V^{\text{eff}}_{\bm{k}\bm{k'}}\rangle=-\big[1-g^{n}_{\mu\nu}(\bm{k})(k_{\mu}k_{\nu}+\langle k_{\mu}'k_{\nu}'\rangle)\big]V,
    \label{eq:Veff-avant-simpl}
\end{equation}

with $\bm{k},\bm{k'}\in\mathcal{D}$. From Eq.~(\ref{eq:Veff-expand}) and the positive semidefiniteness of the quantum metric \cite{Provost1980}, we can then indeed expect the quantum metric to weaken the pairing interaction and the subsequent properties of the superconducting phase. We show in appendix \ref{annexe:Cooper} that the averaged effective interaction for two-dimensional massive Dirac fermions is given by
\begin{equation}
        \frac{\langle V^{\text{eff}}_{\bm{k}\bm{k'}}\rangle}{-V}=1-\frac{\lambdabar_c^2k^2+\lambdabar_c^2\langle k'{}^{2}\rangle(1+\lambdabar_c^2k^2/2)}{4(1+\lambdabar_c^2k^2)^2},
\end{equation}
 where $\lambdabar_c=\hbar v/\abs{\Delta}$ is the reduced Compton wavelength, with $v$ the Dirac velocity and $\Delta$ the band gap. Since we are typically in a regime where the Debye energy is much smaller than the band gap and Fermi energy, we may approximately consider that the pairing only happens in the Fermi surface, such that $\lambdabar_c^2\langle k'^2\rangle=\lambdabar_c^2k^2=\lambdabar_c^2k_F^2$ with $k_F$ the Fermi momentum, and thus
\begin{equation}
    \langle V^{\text{eff}}_{\bm{k}\bm{k'}}\rangle=-\bigg(1-\frac{\lambdabar_c^2k_F^2(4+\lambdabar_c^2k_F^2)}{8(1+\lambdabar_c^2k_F^2)^2}\bigg)V\equiv-V_{\text{eff}}
    \label{eq:Veff/V-MassiveDirac}
\end{equation}

The quantum metric therefore screens the pairing interaction, as a function of $\lambdabar_ck_F$. The behavior of $V_{\text{eff}}$ as a function of $\lambdabar_ck_F$ may be physically interpreted by using the fact that $\lambdabar_c$ is the length associated to the non-locality of massive Dirac fermions, while $\sqrt{2}k_F^{-1}$ is their mean separation length. In the low-doping regime where $\lambdabar_ck_F\ll1$, the non-locality of the massive Dirac fermions happens at a much smaller length than their separation. The non-locality is thus effectively weak, and the quantum metric-driven screening is also expected to be weak. Indeed, in this regime we find $V_{\text{eff}}\simeq(1-\lambdabar_c^2k_F^2/2)V$. In the intermediate regime where $\lambdabar_ck_F\sim1$, the screening can be shown to be maximal at $\lambdabar_ck_F=\sqrt{2}$, i.e. exactly when the non-locality happens on the same scale as the interparticle separation. In this case, the effective interaction is approximately $17\%$ lower than its bare value. Finally, in the high-doping regime where $\lambdabar_ck_F\gg1$ and the non-locality is much stronger than the interparticle separation, the physical interpretation is less clear. Nonetheless, one may argue that in this regime the Fermi level is much higher than the band gap such that the Cooper pairing happens on an energy scale far from the valence band. The quantum geometry, and the subsequent screening, would therefore be weak and ultimately vanish. However we find $V_{\text{eff}}=(7/8-1/(4\lambdabar_c^2k_F^2))V$, such that the screening persists even in the $\lambdabar_ck_F\to\infty$ limit. This counterintuitive fact may be due to the phenomenological character of the derivation we have used, as seen in Sec. \ref{Ssec:Darwin-1corps}.  

The binding energy of the Cooper pair, given by $E_{B}=-(E-2\epsilon_F)$, is then found by solving Eq.~(\ref{eq:self-cons}) in the same manner as conventional BCS theory. This yields
\begin{equation}
    E_B=2\hbar\omega_De^{-2/\lambda_{\text{eff}}},
\end{equation}
with $\lambda_{\text{eff}}=\rho(\epsilon_F)V_{\text{eff}}$ the effective BCS coupling constant, and $\rho(\epsilon_F)$ the density of states at the Fermi level. The expression of the binding energy only differs from its conventional BCS expression by the change from the bare BCS coupling constant $\lambda=\rho(\epsilon_F)V$ to the effective one. More generally one can define an effective BCS Hamiltonian using Eq.~(\ref{eq:Veff}), as done in Ref. \cite{Simon2022}. Studying the mean-field theory of the effective BCS Hamiltonian with the same approximations we made to obtain Eq.~(\ref{eq:Veff/V-MassiveDirac}) also results in the same observation. The critical temperature and superconducting gap then differ from their conventional BCS expression by their dependence on the effective coupling constant. Since the effective coupling constant has an additional dependence on the Fermi momentum, the deviation of the critical temperature from the original BCS value will also depend on $k_F$. This fact could be leveraged to experimentally probe the effect of the Darwin term, by doping (e.g. via a gate voltage) a superconductor.

\section*{Conclusion}

In this work, we have investigated aspects of the relation between the quantum geometry of the normal state and superconductivity. We started by relating band theory to multiparameter quantum estimation theory. The definition of the Bloch Hamiltonian then appears as a parametrization process, and we then showed that the quantum geometric tensor can be written as the covariance matrix in the Bloch state of the position operator. This therefore establishes a momentum-resolved relation between quantum geometry and the quantum fluctuations of the position operator. The quantum metric then quantifies the typical non-locality of the Bloch fermion while the Berry curvature quantifies its minimal non-locality. This non-locality is a consequence of the emergent character of the Bloch fermion.

Upon these results, we then investigated the interpretation of the superfluid weight. The fundamental idea is that Cooper pairing happens between the charge carriers of the normal state, which are the Bloch fermions. On the one hand, the center-of-mass motion of the Bloch fermions drives the conventional supercurrent. On the other hand, the non-locality and zero-point motion of the Bloch fermions drives the geometric supercurrent, which is independent of the center-of-mass motion. Therefore even though the Bloch fermions are immobile as a whole in flat band systems, their non-locality (and therefore emergence) allows them to keep moving in a quantum mechanical sense. We thus argue that the geometric supercurrent is a property of the superconducting phase that is inherited from the normal state. We then rederived and interpreted the geometric and topological bounds on the geometric superfluid weight, from the fact that the Berry curvature (and Chern number) protects a minimal amount of zero-point motion.

Finally, in analogy to elementary electrons in relativistic quantum mechanics, we derived the emergent Darwin term associated to the non-locality of Bloch fermions, when the latter is subject to an external potential. By investigating a generic one-body problem, we showed that this Darwin term is driven by the quantum metric. We then derived the form of the Darwin term in a generic two-body problem, and specified it for a pairing potential between time-reversed partners. We then considered the associated Cooper problem, and derived the form of the effective pairing potential, where the NS quantum metric is anticipated to screen the pairing interaction. In the example of two-dimensional massive Dirac fermions the pairing interaction is indeed screened. This screening is strongest when the non-locality of the massive Dirac fermions happens on the same scale as their typical separation. The fundamental properties of the superconducting phase, such as the binding energy, critical temperature and superconducting gap, then differ from their conventional BCS value by the replacement of the BCS coupling constant by an effective one. The dependence of these results on the Fermi momentum could be leveraged to experimentally probe the screening of the pairing interaction from the emergent Darwin term.

The non-locality of the Bloch fermions, and their emergence, thus appears to be central in understanding the relation between the NS quantum geometry and superconductivity. It is the driver of an ambivalent relationship which can on one side drive a supercurrent but on the other side screen the associated interaction. While we have discussed the case of superconductivity, other correlated phases such as charge density waves could similarly be affected.

Lastly, the non-trivial relationship between the NS quantum geometry and superconductivity highlights the complexity of topological superconductivity and the composite character of the Cooper pairs \cite{Daido2024}.

\section*{Acknowledgements}
The author thanks Frédéric Piéchon for a discussion that motivated this work. The author is also indebted to Mark O. Goerbig for several stimulating discussions and constructive comments on a previous version of the manuscript. The author also thanks Quentin Marsal and Päivi Törmä for helpful comments. This work was supported by the French National Research Agency (projects TWISTGRAPH and QMAHT) under Grants No. ANR-21-CE47-0018 and ANR-22-CE30-0032, respectively.

\bibliography{Biblio1}

\begin{thebibliography}{65}%
\makeatletter
\providecommand \@ifxundefined [1]{%
 \@ifx{#1\undefined}
}%
\providecommand \@ifnum [1]{%
 \ifnum #1\expandafter \@firstoftwo
 \else \expandafter \@secondoftwo
 \fi
}%
\providecommand \@ifx [1]{%
 \ifx #1\expandafter \@firstoftwo
 \else \expandafter \@secondoftwo
 \fi
}%
\providecommand \natexlab [1]{#1}%
\providecommand \enquote  [1]{``#1''}%
\providecommand \bibnamefont  [1]{#1}%
\providecommand \bibfnamefont [1]{#1}%
\providecommand \citenamefont [1]{#1}%
\providecommand \href@noop [0]{\@secondoftwo}%
\providecommand \href [0]{\begingroup \@sanitize@url \@href}%
\providecommand \@href[1]{\@@startlink{#1}\@@href}%
\providecommand \@@href[1]{\endgroup#1\@@endlink}%
\providecommand \@sanitize@url [0]{\catcode `\\12\catcode `\$12\catcode `\&12\catcode `\#12\catcode `\^12\catcode `\_12\catcode `\%12\relax}%
\providecommand \@@startlink[1]{}%
\providecommand \@@endlink[0]{}%
\providecommand \url  [0]{\begingroup\@sanitize@url \@url }%
\providecommand \@url [1]{\endgroup\@href {#1}{\urlprefix }}%
\providecommand \urlprefix  [0]{URL }%
\providecommand \Eprint [0]{\href }%
\providecommand \doibase [0]{https://doi.org/}%
\providecommand \selectlanguage [0]{\@gobble}%
\providecommand \bibinfo  [0]{\@secondoftwo}%
\providecommand \bibfield  [0]{\@secondoftwo}%
\providecommand \translation [1]{[#1]}%
\providecommand \BibitemOpen [0]{}%
\providecommand \bibitemStop [0]{}%
\providecommand \bibitemNoStop [0]{.\EOS\space}%
\providecommand \EOS [0]{\spacefactor3000\relax}%
\providecommand \BibitemShut  [1]{\csname bibitem#1\endcsname}%
\let\auto@bib@innerbib\@empty
\bibitem [{\citenamefont {Cayssol}\ and\ \citenamefont {Fuchs}(2021)}]{Cayssol2021}%
  \BibitemOpen
  \bibfield  {author} {\bibinfo {author} {\bibfnamefont {J.}~\bibnamefont {Cayssol}}\ and\ \bibinfo {author} {\bibfnamefont {J.~N.}\ \bibnamefont {Fuchs}},\ }\bibfield  {title} {\bibinfo {title} {Topological and geometrical aspects of band theory},\ }\href {https://doi.org/10.1088/2515-7639/abf0b5} {\bibfield  {journal} {\bibinfo  {journal} {Journal of Physics: Materials}\ }\textbf {\bibinfo {volume} {4}},\ \bibinfo {pages} {034007} (\bibinfo {year} {2021})}\BibitemShut {NoStop}%
\bibitem [{\citenamefont {Bradlyn}\ \emph {et~al.}(2016)\citenamefont {Bradlyn}, \citenamefont {Cano}, \citenamefont {Wang}, \citenamefont {Vergniory}, \citenamefont {Felser}, \citenamefont {Cava},\ and\ \citenamefont {Bernevig}}]{Bradlyn2016}%
  \BibitemOpen
  \bibfield  {author} {\bibinfo {author} {\bibfnamefont {B.}~\bibnamefont {Bradlyn}}, \bibinfo {author} {\bibfnamefont {J.}~\bibnamefont {Cano}}, \bibinfo {author} {\bibfnamefont {Z.}~\bibnamefont {Wang}}, \bibinfo {author} {\bibfnamefont {M.~G.}\ \bibnamefont {Vergniory}}, \bibinfo {author} {\bibfnamefont {C.}~\bibnamefont {Felser}}, \bibinfo {author} {\bibfnamefont {R.~J.}\ \bibnamefont {Cava}},\ and\ \bibinfo {author} {\bibfnamefont {B.~A.}\ \bibnamefont {Bernevig}},\ }\bibfield  {title} {\bibinfo {title} {Beyond dirac and weyl fermions: Unconventional quasiparticles in conventional crystals},\ }\bibfield  {journal} {\bibinfo  {journal} {Science}\ }\textbf {\bibinfo {volume} {353}},\ \href {https://doi.org/10.1126/science.aaf5037} {10.1126/science.aaf5037} (\bibinfo {year} {2016})\BibitemShut {NoStop}%
\bibitem [{\citenamefont {Qi}\ and\ \citenamefont {Zhang}(2011)}]{Qi2011}%
  \BibitemOpen
  \bibfield  {author} {\bibinfo {author} {\bibfnamefont {X.-L.}\ \bibnamefont {Qi}}\ and\ \bibinfo {author} {\bibfnamefont {S.-C.}\ \bibnamefont {Zhang}},\ }\bibfield  {title} {\bibinfo {title} {Topological insulators and superconductors},\ }\href {https://doi.org/10.1103/revmodphys.83.1057} {\bibfield  {journal} {\bibinfo  {journal} {Reviews of Modern Physics}\ }\textbf {\bibinfo {volume} {83}},\ \bibinfo {pages} {1057–1110} (\bibinfo {year} {2011})}\BibitemShut {NoStop}%
\bibitem [{\citenamefont {T\"orm\"a}(2023)}]{Törma2023}%
  \BibitemOpen
  \bibfield  {author} {\bibinfo {author} {\bibfnamefont {P.}~\bibnamefont {T\"orm\"a}},\ }\bibfield  {title} {\bibinfo {title} {Essay: Where can quantum geometry lead us?},\ }\href {https://doi.org/10.1103/PhysRevLett.131.240001} {\bibfield  {journal} {\bibinfo  {journal} {Phys. Rev. Lett.}\ }\textbf {\bibinfo {volume} {131}},\ \bibinfo {pages} {240001} (\bibinfo {year} {2023})}\BibitemShut {NoStop}%
\bibitem [{\citenamefont {Liu}\ \emph {et~al.}(2024)\citenamefont {Liu}, \citenamefont {Qiang}, \citenamefont {Lu},\ and\ \citenamefont {Xie}}]{Liu2024}%
  \BibitemOpen
  \bibfield  {author} {\bibinfo {author} {\bibfnamefont {T.}~\bibnamefont {Liu}}, \bibinfo {author} {\bibfnamefont {X.-B.}\ \bibnamefont {Qiang}}, \bibinfo {author} {\bibfnamefont {H.-Z.}\ \bibnamefont {Lu}},\ and\ \bibinfo {author} {\bibfnamefont {X.~C.}\ \bibnamefont {Xie}},\ }\bibfield  {title} {\bibinfo {title} {Quantum geometry in condensed matter},\ }\bibfield  {journal} {\bibinfo  {journal} {National Science Review}\ }\textbf {\bibinfo {volume} {12}},\ \href {https://doi.org/10.1093/nsr/nwae334} {10.1093/nsr/nwae334} (\bibinfo {year} {2024})\BibitemShut {NoStop}%
\bibitem [{\citenamefont {Yu}\ \emph {et~al.}(2025)\citenamefont {Yu}, \citenamefont {Bernevig}, \citenamefont {Queiroz}, \citenamefont {Rossi}, \citenamefont {T\"{o}rm\"{a}},\ and\ \citenamefont {Yang}}]{Yu2025}%
  \BibitemOpen
  \bibfield  {author} {\bibinfo {author} {\bibfnamefont {J.}~\bibnamefont {Yu}}, \bibinfo {author} {\bibfnamefont {B.~A.}\ \bibnamefont {Bernevig}}, \bibinfo {author} {\bibfnamefont {R.}~\bibnamefont {Queiroz}}, \bibinfo {author} {\bibfnamefont {E.}~\bibnamefont {Rossi}}, \bibinfo {author} {\bibfnamefont {P.}~\bibnamefont {T\"{o}rm\"{a}}},\ and\ \bibinfo {author} {\bibfnamefont {B.-J.}\ \bibnamefont {Yang}},\ }\href {https://doi.org/10.48550/ARXIV.2501.00098} {\bibinfo {title} {Quantum geometry in quantum materials}} (\bibinfo {year} {2025})\BibitemShut {NoStop}%
\bibitem [{\citenamefont {Roy}(2014)}]{Roy2014}%
  \BibitemOpen
  \bibfield  {author} {\bibinfo {author} {\bibfnamefont {R.}~\bibnamefont {Roy}},\ }\bibfield  {title} {\bibinfo {title} {Band geometry of fractional topological insulators},\ }\href {https://doi.org/10.1103/PhysRevB.90.165139} {\bibfield  {journal} {\bibinfo  {journal} {Phys. Rev. B}\ }\textbf {\bibinfo {volume} {90}},\ \bibinfo {pages} {165139} (\bibinfo {year} {2014})}\BibitemShut {NoStop}%
\bibitem [{\citenamefont {Claassen}\ \emph {et~al.}(2015)\citenamefont {Claassen}, \citenamefont {Lee}, \citenamefont {Thomale}, \citenamefont {Qi},\ and\ \citenamefont {Devereaux}}]{Claassen2015}%
  \BibitemOpen
  \bibfield  {author} {\bibinfo {author} {\bibfnamefont {M.}~\bibnamefont {Claassen}}, \bibinfo {author} {\bibfnamefont {C.~H.}\ \bibnamefont {Lee}}, \bibinfo {author} {\bibfnamefont {R.}~\bibnamefont {Thomale}}, \bibinfo {author} {\bibfnamefont {X.-L.}\ \bibnamefont {Qi}},\ and\ \bibinfo {author} {\bibfnamefont {T.~P.}\ \bibnamefont {Devereaux}},\ }\bibfield  {title} {\bibinfo {title} {Position-momentum duality and fractional quantum hall effect in chern insulators},\ }\bibfield  {journal} {\bibinfo  {journal} {Physical Review Letters}\ }\textbf {\bibinfo {volume} {114}},\ \href {https://doi.org/10.1103/physrevlett.114.236802} {10.1103/physrevlett.114.236802} (\bibinfo {year} {2015})\BibitemShut {NoStop}%
\bibitem [{\citenamefont {Srivastava}\ and\ \citenamefont {Imamo\ifmmode~\breve{g}\else \u{g}\fi{}lu}(2015)}]{Srivastava2015}%
  \BibitemOpen
  \bibfield  {author} {\bibinfo {author} {\bibfnamefont {A.}~\bibnamefont {Srivastava}}\ and\ \bibinfo {author} {\bibfnamefont {A.~m.~c.}\ \bibnamefont {Imamo\ifmmode~\breve{g}\else \u{g}\fi{}lu}},\ }\bibfield  {title} {\bibinfo {title} {Signatures of bloch-band geometry on excitons: Nonhydrogenic spectra in transition-metal dichalcogenides},\ }\href {https://doi.org/10.1103/PhysRevLett.115.166802} {\bibfield  {journal} {\bibinfo  {journal} {Phys. Rev. Lett.}\ }\textbf {\bibinfo {volume} {115}},\ \bibinfo {pages} {166802} (\bibinfo {year} {2015})}\BibitemShut {NoStop}%
\bibitem [{\citenamefont {Zhou}\ \emph {et~al.}(2015)\citenamefont {Zhou}, \citenamefont {Shan}, \citenamefont {Yao},\ and\ \citenamefont {Xiao}}]{Zhou2015}%
  \BibitemOpen
  \bibfield  {author} {\bibinfo {author} {\bibfnamefont {J.}~\bibnamefont {Zhou}}, \bibinfo {author} {\bibfnamefont {W.-Y.}\ \bibnamefont {Shan}}, \bibinfo {author} {\bibfnamefont {W.}~\bibnamefont {Yao}},\ and\ \bibinfo {author} {\bibfnamefont {D.}~\bibnamefont {Xiao}},\ }\bibfield  {title} {\bibinfo {title} {Berry phase modification to the energy spectrum of excitons},\ }\bibfield  {journal} {\bibinfo  {journal} {Physical Review Letters}\ }\textbf {\bibinfo {volume} {115}},\ \href {https://doi.org/10.1103/physrevlett.115.166803} {10.1103/physrevlett.115.166803} (\bibinfo {year} {2015})\BibitemShut {NoStop}%
\bibitem [{\citenamefont {Hichri}\ \emph {et~al.}(2019)\citenamefont {Hichri}, \citenamefont {Jaziri},\ and\ \citenamefont {Goerbig}}]{Hichri2019}%
  \BibitemOpen
  \bibfield  {author} {\bibinfo {author} {\bibfnamefont {A.}~\bibnamefont {Hichri}}, \bibinfo {author} {\bibfnamefont {S.}~\bibnamefont {Jaziri}},\ and\ \bibinfo {author} {\bibfnamefont {M.~O.}\ \bibnamefont {Goerbig}},\ }\bibfield  {title} {\bibinfo {title} {Charged excitons in two-dimensional transition metal dichalcogenides: Semiclassical calculation of berry curvature effects},\ }\href {https://doi.org/10.1103/PhysRevB.100.115426} {\bibfield  {journal} {\bibinfo  {journal} {Phys. Rev. B}\ }\textbf {\bibinfo {volume} {100}},\ \bibinfo {pages} {115426} (\bibinfo {year} {2019})}\BibitemShut {NoStop}%
\bibitem [{\citenamefont {Cao}\ \emph {et~al.}(2021)\citenamefont {Cao}, \citenamefont {Fertig},\ and\ \citenamefont {Brey}}]{Cao2021}%
  \BibitemOpen
  \bibfield  {author} {\bibinfo {author} {\bibfnamefont {J.}~\bibnamefont {Cao}}, \bibinfo {author} {\bibfnamefont {H.~A.}\ \bibnamefont {Fertig}},\ and\ \bibinfo {author} {\bibfnamefont {L.}~\bibnamefont {Brey}},\ }\bibfield  {title} {\bibinfo {title} {Quantum geometric exciton drift velocity},\ }\bibfield  {journal} {\bibinfo  {journal} {Physical Review B}\ }\textbf {\bibinfo {volume} {103}},\ \href {https://doi.org/10.1103/physrevb.103.115422} {10.1103/physrevb.103.115422} (\bibinfo {year} {2021})\BibitemShut {NoStop}%
\bibitem [{\citenamefont {Peotta}\ and\ \citenamefont {T\"{o}rm\"{a}}(2015)}]{Peotta2015}%
  \BibitemOpen
  \bibfield  {author} {\bibinfo {author} {\bibfnamefont {S.}~\bibnamefont {Peotta}}\ and\ \bibinfo {author} {\bibfnamefont {P.}~\bibnamefont {T\"{o}rm\"{a}}},\ }\bibfield  {title} {\bibinfo {title} {Superfluidity in topologically nontrivial flat bands},\ }\bibfield  {journal} {\bibinfo  {journal} {Nature Communications}\ }\textbf {\bibinfo {volume} {6}},\ \href {https://doi.org/10.1038/ncomms9944} {10.1038/ncomms9944} (\bibinfo {year} {2015})\BibitemShut {NoStop}%
\bibitem [{\citenamefont {Rossi}(2021)}]{Rossi2021}%
  \BibitemOpen
  \bibfield  {author} {\bibinfo {author} {\bibfnamefont {E.}~\bibnamefont {Rossi}},\ }\bibfield  {title} {\bibinfo {title} {Quantum metric and correlated states in two-dimensional systems},\ }\href {https://doi.org/10.1016/j.cossms.2021.100952} {\bibfield  {journal} {\bibinfo  {journal} {Current Opinion in Solid State and Materials Science}\ }\textbf {\bibinfo {volume} {25}},\ \bibinfo {pages} {100952} (\bibinfo {year} {2021})}\BibitemShut {NoStop}%
\bibitem [{\citenamefont {T\"{o}rm\"{a}}\ \emph {et~al.}(2022)\citenamefont {T\"{o}rm\"{a}}, \citenamefont {Peotta},\ and\ \citenamefont {Bernevig}}]{Trm2022}%
  \BibitemOpen
  \bibfield  {author} {\bibinfo {author} {\bibfnamefont {P.}~\bibnamefont {T\"{o}rm\"{a}}}, \bibinfo {author} {\bibfnamefont {S.}~\bibnamefont {Peotta}},\ and\ \bibinfo {author} {\bibfnamefont {B.~A.}\ \bibnamefont {Bernevig}},\ }\bibfield  {title} {\bibinfo {title} {Superconductivity, superfluidity and quantum geometry in twisted multilayer systems},\ }\href {https://doi.org/10.1038/s42254-022-00466-y} {\bibfield  {journal} {\bibinfo  {journal} {Nature Reviews Physics}\ }\textbf {\bibinfo {volume} {4}},\ \bibinfo {pages} {528–542} (\bibinfo {year} {2022})}\BibitemShut {NoStop}%
\bibitem [{\citenamefont {Simon}\ \emph {et~al.}(2022)\citenamefont {Simon}, \citenamefont {Gabay}, \citenamefont {Goerbig},\ and\ \citenamefont {Pagot}}]{Simon2022}%
  \BibitemOpen
  \bibfield  {author} {\bibinfo {author} {\bibfnamefont {F.}~\bibnamefont {Simon}}, \bibinfo {author} {\bibfnamefont {M.}~\bibnamefont {Gabay}}, \bibinfo {author} {\bibfnamefont {M.~O.}\ \bibnamefont {Goerbig}},\ and\ \bibinfo {author} {\bibfnamefont {L.}~\bibnamefont {Pagot}},\ }\bibfield  {title} {\bibinfo {title} {Role of the berry curvature on bcs-type superconductivity in two-dimensional materials},\ }\href {https://doi.org/10.1103/PhysRevB.106.214512} {\bibfield  {journal} {\bibinfo  {journal} {Phys. Rev. B}\ }\textbf {\bibinfo {volume} {106}},\ \bibinfo {pages} {214512} (\bibinfo {year} {2022})}\BibitemShut {NoStop}%
\bibitem [{\citenamefont {Kitamura}\ \emph {et~al.}(2022{\natexlab{a}})\citenamefont {Kitamura}, \citenamefont {Daido},\ and\ \citenamefont {Yanase}}]{Kitamura2022}%
  \BibitemOpen
  \bibfield  {author} {\bibinfo {author} {\bibfnamefont {T.}~\bibnamefont {Kitamura}}, \bibinfo {author} {\bibfnamefont {A.}~\bibnamefont {Daido}},\ and\ \bibinfo {author} {\bibfnamefont {Y.}~\bibnamefont {Yanase}},\ }\bibfield  {title} {\bibinfo {title} {Quantum geometric effect on fulde-ferrell-larkin-ovchinnikov superconductivity},\ }\href {https://doi.org/10.1103/PhysRevB.106.184507} {\bibfield  {journal} {\bibinfo  {journal} {Phys. Rev. B}\ }\textbf {\bibinfo {volume} {106}},\ \bibinfo {pages} {184507} (\bibinfo {year} {2022}{\natexlab{a}})}\BibitemShut {NoStop}%
\bibitem [{\citenamefont {Iskin}(2023)}]{Iskin2023}%
  \BibitemOpen
  \bibfield  {author} {\bibinfo {author} {\bibfnamefont {M.}~\bibnamefont {Iskin}},\ }\bibfield  {title} {\bibinfo {title} {Extracting quantum-geometric effects from ginzburg-landau theory in a multiband hubbard model},\ }\bibfield  {journal} {\bibinfo  {journal} {Physical Review B}\ }\textbf {\bibinfo {volume} {107}},\ \href {https://doi.org/10.1103/physrevb.107.224505} {10.1103/physrevb.107.224505} (\bibinfo {year} {2023})\BibitemShut {NoStop}%
\bibitem [{\citenamefont {Porlles}\ and\ \citenamefont {Chen}(2023)}]{Porlles2023}%
  \BibitemOpen
  \bibfield  {author} {\bibinfo {author} {\bibfnamefont {D.}~\bibnamefont {Porlles}}\ and\ \bibinfo {author} {\bibfnamefont {W.}~\bibnamefont {Chen}},\ }\bibfield  {title} {\bibinfo {title} {Quantum geometry of singlet superconductors},\ }\bibfield  {journal} {\bibinfo  {journal} {Physical Review B}\ }\textbf {\bibinfo {volume} {108}},\ \href {https://doi.org/10.1103/physrevb.108.094508} {10.1103/physrevb.108.094508} (\bibinfo {year} {2023})\BibitemShut {NoStop}%
\bibitem [{\citenamefont {Kitamura}\ \emph {et~al.}(2023)\citenamefont {Kitamura}, \citenamefont {Kanasugi}, \citenamefont {Chazono},\ and\ \citenamefont {Yanase}}]{Kitamura2023}%
  \BibitemOpen
  \bibfield  {author} {\bibinfo {author} {\bibfnamefont {T.}~\bibnamefont {Kitamura}}, \bibinfo {author} {\bibfnamefont {S.}~\bibnamefont {Kanasugi}}, \bibinfo {author} {\bibfnamefont {M.}~\bibnamefont {Chazono}},\ and\ \bibinfo {author} {\bibfnamefont {Y.}~\bibnamefont {Yanase}},\ }\bibfield  {title} {\bibinfo {title} {Quantum geometry induced anapole superconductivity},\ }\href {https://doi.org/10.1103/PhysRevB.107.214513} {\bibfield  {journal} {\bibinfo  {journal} {Phys. Rev. B}\ }\textbf {\bibinfo {volume} {107}},\ \bibinfo {pages} {214513} (\bibinfo {year} {2023})}\BibitemShut {NoStop}%
\bibitem [{\citenamefont {Chen}\ and\ \citenamefont {Law}(2024)}]{Chen2024}%
  \BibitemOpen
  \bibfield  {author} {\bibinfo {author} {\bibfnamefont {S.~A.}\ \bibnamefont {Chen}}\ and\ \bibinfo {author} {\bibfnamefont {K.~T.}\ \bibnamefont {Law}},\ }\bibfield  {title} {\bibinfo {title} {Ginzburg-landau theory of flat-band superconductors with quantum metric},\ }\bibfield  {journal} {\bibinfo  {journal} {Physical Review Letters}\ }\textbf {\bibinfo {volume} {132}},\ \href {https://doi.org/10.1103/physrevlett.132.026002} {10.1103/physrevlett.132.026002} (\bibinfo {year} {2024})\BibitemShut {NoStop}%
\bibitem [{\citenamefont {Kitamura}\ \emph {et~al.}(2024)\citenamefont {Kitamura}, \citenamefont {Daido},\ and\ \citenamefont {Yanase}}]{Kitamura2024}%
  \BibitemOpen
  \bibfield  {author} {\bibinfo {author} {\bibfnamefont {T.}~\bibnamefont {Kitamura}}, \bibinfo {author} {\bibfnamefont {A.}~\bibnamefont {Daido}},\ and\ \bibinfo {author} {\bibfnamefont {Y.}~\bibnamefont {Yanase}},\ }\bibfield  {title} {\bibinfo {title} {Spin-triplet superconductivity from quantum-geometry-induced ferromagnetic fluctuation},\ }\href {https://doi.org/10.1103/PhysRevLett.132.036001} {\bibfield  {journal} {\bibinfo  {journal} {Phys. Rev. Lett.}\ }\textbf {\bibinfo {volume} {132}},\ \bibinfo {pages} {036001} (\bibinfo {year} {2024})}\BibitemShut {NoStop}%
\bibitem [{\citenamefont {Daido}\ \emph {et~al.}(2024)\citenamefont {Daido}, \citenamefont {Kitamura},\ and\ \citenamefont {Yanase}}]{Daido2024}%
  \BibitemOpen
  \bibfield  {author} {\bibinfo {author} {\bibfnamefont {A.}~\bibnamefont {Daido}}, \bibinfo {author} {\bibfnamefont {T.}~\bibnamefont {Kitamura}},\ and\ \bibinfo {author} {\bibfnamefont {Y.}~\bibnamefont {Yanase}},\ }\bibfield  {title} {\bibinfo {title} {Quantum geometry encoded to pair potentials},\ }\bibfield  {journal} {\bibinfo  {journal} {Physical Review B}\ }\textbf {\bibinfo {volume} {110}},\ \href {https://doi.org/10.1103/physrevb.110.094505} {10.1103/physrevb.110.094505} (\bibinfo {year} {2024})\BibitemShut {NoStop}%
\bibitem [{\citenamefont {Yu}\ \emph {et~al.}(2024)\citenamefont {Yu}, \citenamefont {Ciccarino}, \citenamefont {Bianco}, \citenamefont {Errea}, \citenamefont {Narang},\ and\ \citenamefont {Bernevig}}]{Yu2024}%
  \BibitemOpen
  \bibfield  {author} {\bibinfo {author} {\bibfnamefont {J.}~\bibnamefont {Yu}}, \bibinfo {author} {\bibfnamefont {C.~J.}\ \bibnamefont {Ciccarino}}, \bibinfo {author} {\bibfnamefont {R.}~\bibnamefont {Bianco}}, \bibinfo {author} {\bibfnamefont {I.}~\bibnamefont {Errea}}, \bibinfo {author} {\bibfnamefont {P.}~\bibnamefont {Narang}},\ and\ \bibinfo {author} {\bibfnamefont {B.~A.}\ \bibnamefont {Bernevig}},\ }\bibfield  {title} {\bibinfo {title} {Non-trivial quantum geometry and the strength of electron–phonon coupling},\ }\bibfield  {journal} {\bibinfo  {journal} {Nature Physics}\ }\href {https://doi.org/10.1038/s41567-024-02486-0} {10.1038/s41567-024-02486-0} (\bibinfo {year} {2024})\BibitemShut {NoStop}%
\bibitem [{\citenamefont {Wang}\ \emph {et~al.}(2024)\citenamefont {Wang}, \citenamefont {Gao},\ and\ \citenamefont {Yang}}]{Wang2024}%
  \BibitemOpen
  \bibfield  {author} {\bibinfo {author} {\bibfnamefont {Y.-Q.}\ \bibnamefont {Wang}}, \bibinfo {author} {\bibfnamefont {Z.-Q.}\ \bibnamefont {Gao}},\ and\ \bibinfo {author} {\bibfnamefont {H.}~\bibnamefont {Yang}},\ }\href {https://doi.org/10.48550/ARXIV.2410.05384} {\bibinfo {title} {Chiral superconductivity from parent chern band and its non-abelian generalization}} (\bibinfo {year} {2024})\BibitemShut {NoStop}%
\bibitem [{\citenamefont {Hu}\ \emph {et~al.}(2025)\citenamefont {Hu}, \citenamefont {Chen},\ and\ \citenamefont {Law}}]{Hu2025b}%
  \BibitemOpen
  \bibfield  {author} {\bibinfo {author} {\bibfnamefont {J.-X.}\ \bibnamefont {Hu}}, \bibinfo {author} {\bibfnamefont {S.~A.}\ \bibnamefont {Chen}},\ and\ \bibinfo {author} {\bibfnamefont {K.~T.}\ \bibnamefont {Law}},\ }\bibfield  {title} {\bibinfo {title} {Anomalous coherence length in superconductors with quantum metric},\ }\bibfield  {journal} {\bibinfo  {journal} {Communications Physics}\ }\textbf {\bibinfo {volume} {8}},\ \href {https://doi.org/10.1038/s42005-024-01930-0} {10.1038/s42005-024-01930-0} (\bibinfo {year} {2025})\BibitemShut {NoStop}%
\bibitem [{\citenamefont {Li}\ \emph {et~al.}(2025)\citenamefont {Li}, \citenamefont {Zhang},\ and\ \citenamefont {Hu}}]{Li2025}%
  \BibitemOpen
  \bibfield  {author} {\bibinfo {author} {\bibfnamefont {C.}~\bibnamefont {Li}}, \bibinfo {author} {\bibfnamefont {F.-C.}\ \bibnamefont {Zhang}},\ and\ \bibinfo {author} {\bibfnamefont {L.-H.}\ \bibnamefont {Hu}},\ }\href {https://doi.org/10.48550/arxiv.2505.01682} {\bibinfo {title} {Vortex states and coherence lengths in flat-band superconductors}} (\bibinfo {year} {2025})\BibitemShut {NoStop}%
\bibitem [{\citenamefont {Thumin}\ and\ \citenamefont {Bouzerar}(2025)}]{Thumin2025}%
  \BibitemOpen
  \bibfield  {author} {\bibinfo {author} {\bibfnamefont {M.}~\bibnamefont {Thumin}}\ and\ \bibinfo {author} {\bibfnamefont {G.}~\bibnamefont {Bouzerar}},\ }\bibfield  {title} {\bibinfo {title} {Correlation functions and characteristic lengthscales in flat band superconductors},\ }\bibfield  {journal} {\bibinfo  {journal} {SciPost Physics}\ }\textbf {\bibinfo {volume} {18}},\ \href {https://doi.org/10.21468/scipostphys.18.1.025} {10.21468/scipostphys.18.1.025} (\bibinfo {year} {2025})\BibitemShut {NoStop}%
\bibitem [{\citenamefont {Marzari}\ and\ \citenamefont {Vanderbilt}(1997)}]{Marzari1997}%
  \BibitemOpen
  \bibfield  {author} {\bibinfo {author} {\bibfnamefont {N.}~\bibnamefont {Marzari}}\ and\ \bibinfo {author} {\bibfnamefont {D.}~\bibnamefont {Vanderbilt}},\ }\bibfield  {title} {\bibinfo {title} {Maximally localized generalized wannier functions for composite energy bands},\ }\href {https://doi.org/10.1103/PhysRevB.56.12847} {\bibfield  {journal} {\bibinfo  {journal} {Phys. Rev. B}\ }\textbf {\bibinfo {volume} {56}},\ \bibinfo {pages} {12847} (\bibinfo {year} {1997})}\BibitemShut {NoStop}%
\bibitem [{\citenamefont {Resta}(2011)}]{Resta2011}%
  \BibitemOpen
  \bibfield  {author} {\bibinfo {author} {\bibfnamefont {R.}~\bibnamefont {Resta}},\ }\bibfield  {title} {\bibinfo {title} {The insulating state of matter: a geometrical theory},\ }\href {https://doi.org/10.1140/epjb/e2010-10874-4} {\bibfield  {journal} {\bibinfo  {journal} {The European Physical Journal B}\ }\textbf {\bibinfo {volume} {79}},\ \bibinfo {pages} {121–137} (\bibinfo {year} {2011})}\BibitemShut {NoStop}%
\bibitem [{\citenamefont {Boixo}\ \emph {et~al.}(2007)\citenamefont {Boixo}, \citenamefont {Flammia}, \citenamefont {Caves},\ and\ \citenamefont {Geremia}}]{Boixo2007}%
  \BibitemOpen
  \bibfield  {author} {\bibinfo {author} {\bibfnamefont {S.}~\bibnamefont {Boixo}}, \bibinfo {author} {\bibfnamefont {S.~T.}\ \bibnamefont {Flammia}}, \bibinfo {author} {\bibfnamefont {C.~M.}\ \bibnamefont {Caves}},\ and\ \bibinfo {author} {\bibfnamefont {J.}~\bibnamefont {Geremia}},\ }\bibfield  {title} {\bibinfo {title} {Generalized limits for single-parameter quantum estimation},\ }\bibfield  {journal} {\bibinfo  {journal} {Physical Review Letters}\ }\textbf {\bibinfo {volume} {98}},\ \href {https://doi.org/10.1103/physrevlett.98.090401} {10.1103/physrevlett.98.090401} (\bibinfo {year} {2007})\BibitemShut {NoStop}%
\bibitem [{\citenamefont {Pang}\ and\ \citenamefont {Brun}(2014)}]{Pang2014}%
  \BibitemOpen
  \bibfield  {author} {\bibinfo {author} {\bibfnamefont {S.}~\bibnamefont {Pang}}\ and\ \bibinfo {author} {\bibfnamefont {T.~A.}\ \bibnamefont {Brun}},\ }\bibfield  {title} {\bibinfo {title} {Quantum metrology for a general hamiltonian parameter},\ }\bibfield  {journal} {\bibinfo  {journal} {Physical Review A}\ }\textbf {\bibinfo {volume} {90}},\ \href {https://doi.org/10.1103/physreva.90.022117} {10.1103/physreva.90.022117} (\bibinfo {year} {2014})\BibitemShut {NoStop}%
\bibitem [{\citenamefont {Guo}\ \emph {et~al.}(2016)\citenamefont {Guo}, \citenamefont {Zhong}, \citenamefont {Jing}, \citenamefont {Fu},\ and\ \citenamefont {Wang}}]{Guo2016}%
  \BibitemOpen
  \bibfield  {author} {\bibinfo {author} {\bibfnamefont {W.}~\bibnamefont {Guo}}, \bibinfo {author} {\bibfnamefont {W.}~\bibnamefont {Zhong}}, \bibinfo {author} {\bibfnamefont {X.-X.}\ \bibnamefont {Jing}}, \bibinfo {author} {\bibfnamefont {L.-B.}\ \bibnamefont {Fu}},\ and\ \bibinfo {author} {\bibfnamefont {X.}~\bibnamefont {Wang}},\ }\bibfield  {title} {\bibinfo {title} {Berry curvature as a lower bound for multiparameter estimation},\ }\bibfield  {journal} {\bibinfo  {journal} {Physical Review A}\ }\textbf {\bibinfo {volume} {93}},\ \href {https://doi.org/10.1103/physreva.93.042115} {10.1103/physreva.93.042115} (\bibinfo {year} {2016})\BibitemShut {NoStop}%
\bibitem [{\citenamefont {Ragy}\ \emph {et~al.}(2016)\citenamefont {Ragy}, \citenamefont {Jarzyna},\ and\ \citenamefont {Demkowicz-Dobrzański}}]{Ragy2016}%
  \BibitemOpen
  \bibfield  {author} {\bibinfo {author} {\bibfnamefont {S.}~\bibnamefont {Ragy}}, \bibinfo {author} {\bibfnamefont {M.}~\bibnamefont {Jarzyna}},\ and\ \bibinfo {author} {\bibfnamefont {R.}~\bibnamefont {Demkowicz-Dobrzański}},\ }\bibfield  {title} {\bibinfo {title} {Compatibility in multiparameter quantum metrology},\ }\bibfield  {journal} {\bibinfo  {journal} {Physical Review A}\ }\textbf {\bibinfo {volume} {94}},\ \href {https://doi.org/10.1103/physreva.94.052108} {10.1103/physreva.94.052108} (\bibinfo {year} {2016})\BibitemShut {NoStop}%
\bibitem [{\citenamefont {Liu}\ \emph {et~al.}(2019)\citenamefont {Liu}, \citenamefont {Yuan}, \citenamefont {Lu},\ and\ \citenamefont {Wang}}]{Liu2019}%
  \BibitemOpen
  \bibfield  {author} {\bibinfo {author} {\bibfnamefont {J.}~\bibnamefont {Liu}}, \bibinfo {author} {\bibfnamefont {H.}~\bibnamefont {Yuan}}, \bibinfo {author} {\bibfnamefont {X.-M.}\ \bibnamefont {Lu}},\ and\ \bibinfo {author} {\bibfnamefont {X.}~\bibnamefont {Wang}},\ }\bibfield  {title} {\bibinfo {title} {Quantum fisher information matrix and multiparameter estimation},\ }\href {https://doi.org/10.1088/1751-8121/ab5d4d} {\bibfield  {journal} {\bibinfo  {journal} {Journal of Physics A: Mathematical and Theoretical}\ }\textbf {\bibinfo {volume} {53}},\ \bibinfo {pages} {023001} (\bibinfo {year} {2019})}\BibitemShut {NoStop}%
\bibitem [{\citenamefont {Carollo}\ \emph {et~al.}(2019)\citenamefont {Carollo}, \citenamefont {Spagnolo}, \citenamefont {Dubkov},\ and\ \citenamefont {Valenti}}]{Carollo2019}%
  \BibitemOpen
  \bibfield  {author} {\bibinfo {author} {\bibfnamefont {A.}~\bibnamefont {Carollo}}, \bibinfo {author} {\bibfnamefont {B.}~\bibnamefont {Spagnolo}}, \bibinfo {author} {\bibfnamefont {A.~A.}\ \bibnamefont {Dubkov}},\ and\ \bibinfo {author} {\bibfnamefont {D.}~\bibnamefont {Valenti}},\ }\bibfield  {title} {\bibinfo {title} {On quantumness in multi-parameter quantum estimation},\ }\href {https://doi.org/10.1088/1742-5468/ab3ccb} {\bibfield  {journal} {\bibinfo  {journal} {Journal of Statistical Mechanics: Theory and Experiment}\ }\textbf {\bibinfo {volume} {2019}},\ \bibinfo {pages} {094010} (\bibinfo {year} {2019})}\BibitemShut {NoStop}%
\bibitem [{\citenamefont {Xing}\ and\ \citenamefont {Fu}(2020)}]{Xing2020}%
  \BibitemOpen
  \bibfield  {author} {\bibinfo {author} {\bibfnamefont {H.}~\bibnamefont {Xing}}\ and\ \bibinfo {author} {\bibfnamefont {L.}~\bibnamefont {Fu}},\ }\bibfield  {title} {\bibinfo {title} {Measure of the density of quantum states in information geometry and quantum multiparameter estimation},\ }\bibfield  {journal} {\bibinfo  {journal} {Physical Review A}\ }\textbf {\bibinfo {volume} {102}},\ \href {https://doi.org/10.1103/physreva.102.062613} {10.1103/physreva.102.062613} (\bibinfo {year} {2020})\BibitemShut {NoStop}%
\bibitem [{\citenamefont {Lu}\ and\ \citenamefont {Wang}(2021)}]{Lu2021}%
  \BibitemOpen
  \bibfield  {author} {\bibinfo {author} {\bibfnamefont {X.-M.}\ \bibnamefont {Lu}}\ and\ \bibinfo {author} {\bibfnamefont {X.}~\bibnamefont {Wang}},\ }\bibfield  {title} {\bibinfo {title} {Incorporating heisenberg’s uncertainty principle into quantum multiparameter estimation},\ }\bibfield  {journal} {\bibinfo  {journal} {Physical Review Letters}\ }\textbf {\bibinfo {volume} {126}},\ \href {https://doi.org/10.1103/physrevlett.126.120503} {10.1103/physrevlett.126.120503} (\bibinfo {year} {2021})\BibitemShut {NoStop}%
\bibitem [{\citenamefont {Xia}\ \emph {et~al.}(2023)\citenamefont {Xia}, \citenamefont {Huang}, \citenamefont {Li}, \citenamefont {Wang},\ and\ \citenamefont {Zeng}}]{Xia2023}%
  \BibitemOpen
  \bibfield  {author} {\bibinfo {author} {\bibfnamefont {B.}~\bibnamefont {Xia}}, \bibinfo {author} {\bibfnamefont {J.}~\bibnamefont {Huang}}, \bibinfo {author} {\bibfnamefont {H.}~\bibnamefont {Li}}, \bibinfo {author} {\bibfnamefont {H.}~\bibnamefont {Wang}},\ and\ \bibinfo {author} {\bibfnamefont {G.}~\bibnamefont {Zeng}},\ }\bibfield  {title} {\bibinfo {title} {Toward incompatible quantum limits on multiparameter estimation},\ }\bibfield  {journal} {\bibinfo  {journal} {Nature Communications}\ }\textbf {\bibinfo {volume} {14}},\ \href {https://doi.org/10.1038/s41467-023-36661-3} {10.1038/s41467-023-36661-3} (\bibinfo {year} {2023})\BibitemShut {NoStop}%
\bibitem [{\citenamefont {Lambert}\ and\ \citenamefont {Sørensen}(2023)}]{Lambert2023}%
  \BibitemOpen
  \bibfield  {author} {\bibinfo {author} {\bibfnamefont {J.}~\bibnamefont {Lambert}}\ and\ \bibinfo {author} {\bibfnamefont {E.~S.}\ \bibnamefont {Sørensen}},\ }\bibfield  {title} {\bibinfo {title} {From classical to quantum information geometry: a guide for physicists},\ }\href {https://doi.org/10.1088/1367-2630/aceb14} {\bibfield  {journal} {\bibinfo  {journal} {New Journal of Physics}\ }\textbf {\bibinfo {volume} {25}},\ \bibinfo {pages} {081201} (\bibinfo {year} {2023})}\BibitemShut {NoStop}%
\bibitem [{\citenamefont {Provost}\ and\ \citenamefont {Vallee}(1980)}]{Provost1980}%
  \BibitemOpen
  \bibfield  {author} {\bibinfo {author} {\bibfnamefont {J.~P.}\ \bibnamefont {Provost}}\ and\ \bibinfo {author} {\bibfnamefont {G.}~\bibnamefont {Vallee}},\ }\href@noop {} {\bibfield  {journal} {\bibinfo  {journal} {Communications in Mathematical Physics}\ }\textbf {\bibinfo {volume} {76}},\ \bibinfo {pages} {289–301} (\bibinfo {year} {1980})}\BibitemShut {NoStop}%
\bibitem [{\citenamefont {Komissarov}\ \emph {et~al.}(2024)\citenamefont {Komissarov}, \citenamefont {Holder},\ and\ \citenamefont {Queiroz}}]{Komissarov2024}%
  \BibitemOpen
  \bibfield  {author} {\bibinfo {author} {\bibfnamefont {I.}~\bibnamefont {Komissarov}}, \bibinfo {author} {\bibfnamefont {T.}~\bibnamefont {Holder}},\ and\ \bibinfo {author} {\bibfnamefont {R.}~\bibnamefont {Queiroz}},\ }\bibfield  {title} {\bibinfo {title} {The quantum geometric origin of capacitance in insulators},\ }\bibfield  {journal} {\bibinfo  {journal} {Nature Communications}\ }\textbf {\bibinfo {volume} {15}},\ \href {https://doi.org/10.1038/s41467-024-48808-x} {10.1038/s41467-024-48808-x} (\bibinfo {year} {2024})\BibitemShut {NoStop}%
\bibitem [{\citenamefont {Blount}(1962)}]{Blount1962}%
  \BibitemOpen
  \bibfield  {author} {\bibinfo {author} {\bibfnamefont {E.}~\bibnamefont {Blount}},\ }\bibinfo {title} {Formalisms of band theory},\ in\ \href {https://doi.org/10.1016/s0081-1947(08)60459-2} {\emph {\bibinfo {booktitle} {Solid State Physics}}}\ (\bibinfo  {publisher} {Elsevier},\ \bibinfo {year} {1962})\ p.\ \bibinfo {pages} {305–373}\BibitemShut {NoStop}%
\bibitem [{\citenamefont {Zelevinsky}(2010)}]{Zelevinsky2010}%
  \BibitemOpen
  \bibfield  {author} {\bibinfo {author} {\bibfnamefont {V.}~\bibnamefont {Zelevinsky}},\ }\href@noop {} {\emph {\bibinfo {title} {Quantum physics}}},\ Vol.~\bibinfo {volume} {1}\ (\bibinfo  {publisher} {Wiley-VCH Verlag},\ \bibinfo {address} {Weinheim, Germany},\ \bibinfo {year} {2010})\ p.\ \bibinfo {pages} {551}\BibitemShut {NoStop}%
\bibitem [{\citenamefont {Liang}\ \emph {et~al.}(2017)\citenamefont {Liang}, \citenamefont {Vanhala}, \citenamefont {Peotta}, \citenamefont {Siro}, \citenamefont {Harju},\ and\ \citenamefont {T\"{o}rm\"{a}}}]{Liang2017}%
  \BibitemOpen
  \bibfield  {author} {\bibinfo {author} {\bibfnamefont {L.}~\bibnamefont {Liang}}, \bibinfo {author} {\bibfnamefont {T.~I.}\ \bibnamefont {Vanhala}}, \bibinfo {author} {\bibfnamefont {S.}~\bibnamefont {Peotta}}, \bibinfo {author} {\bibfnamefont {T.}~\bibnamefont {Siro}}, \bibinfo {author} {\bibfnamefont {A.}~\bibnamefont {Harju}},\ and\ \bibinfo {author} {\bibfnamefont {P.}~\bibnamefont {T\"{o}rm\"{a}}},\ }\bibfield  {title} {\bibinfo {title} {Band geometry, berry curvature, and superfluid weight},\ }\bibfield  {journal} {\bibinfo  {journal} {Physical Review B}\ }\textbf {\bibinfo {volume} {95}},\ \href {https://doi.org/10.1103/physrevb.95.024515} {10.1103/physrevb.95.024515} (\bibinfo {year} {2017})\BibitemShut {NoStop}%
\bibitem [{\citenamefont {Chandrasekhar}\ and\ \citenamefont {Einzel}(1993)}]{Chandrasekhar1993}%
  \BibitemOpen
  \bibfield  {author} {\bibinfo {author} {\bibfnamefont {B.~S.}\ \bibnamefont {Chandrasekhar}}\ and\ \bibinfo {author} {\bibfnamefont {D.}~\bibnamefont {Einzel}},\ }\bibfield  {title} {\bibinfo {title} {The superconducting penetration depth from the semiclassical model},\ }\href {https://doi.org/10.1002/andp.19935050604} {\bibfield  {journal} {\bibinfo  {journal} {Annalen der Physik}\ }\textbf {\bibinfo {volume} {505}},\ \bibinfo {pages} {535–546} (\bibinfo {year} {1993})}\BibitemShut {NoStop}%
\bibitem [{\citenamefont {Tian}\ \emph {et~al.}(2023)\citenamefont {Tian}, \citenamefont {Gao}, \citenamefont {Zhang}, \citenamefont {Che}, \citenamefont {Xu}, \citenamefont {Cheung}, \citenamefont {Watanabe}, \citenamefont {Taniguchi}, \citenamefont {Randeria}, \citenamefont {Zhang}, \citenamefont {Lau},\ and\ \citenamefont {Bockrath}}]{Tian2023}%
  \BibitemOpen
  \bibfield  {author} {\bibinfo {author} {\bibfnamefont {H.}~\bibnamefont {Tian}}, \bibinfo {author} {\bibfnamefont {X.}~\bibnamefont {Gao}}, \bibinfo {author} {\bibfnamefont {Y.}~\bibnamefont {Zhang}}, \bibinfo {author} {\bibfnamefont {S.}~\bibnamefont {Che}}, \bibinfo {author} {\bibfnamefont {T.}~\bibnamefont {Xu}}, \bibinfo {author} {\bibfnamefont {P.}~\bibnamefont {Cheung}}, \bibinfo {author} {\bibfnamefont {K.}~\bibnamefont {Watanabe}}, \bibinfo {author} {\bibfnamefont {T.}~\bibnamefont {Taniguchi}}, \bibinfo {author} {\bibfnamefont {M.}~\bibnamefont {Randeria}}, \bibinfo {author} {\bibfnamefont {F.}~\bibnamefont {Zhang}}, \bibinfo {author} {\bibfnamefont {C.~N.}\ \bibnamefont {Lau}},\ and\ \bibinfo {author} {\bibfnamefont {M.~W.}\ \bibnamefont {Bockrath}},\ }\bibfield  {title} {\bibinfo {title} {Evidence for dirac flat band superconductivity enabled by quantum geometry},\ }\href {https://doi.org/10.1038/s41586-022-05576-2} {\bibfield  {journal} {\bibinfo  {journal} {Nature}\ }\textbf {\bibinfo {volume}
  {614}},\ \bibinfo {pages} {440–444} (\bibinfo {year} {2023})}\BibitemShut {NoStop}%
\bibitem [{\citenamefont {Kitamura}\ \emph {et~al.}(2022{\natexlab{b}})\citenamefont {Kitamura}, \citenamefont {Yamashita}, \citenamefont {Ishizuka}, \citenamefont {Daido},\ and\ \citenamefont {Yanase}}]{Kitamura2022b}%
  \BibitemOpen
  \bibfield  {author} {\bibinfo {author} {\bibfnamefont {T.}~\bibnamefont {Kitamura}}, \bibinfo {author} {\bibfnamefont {T.}~\bibnamefont {Yamashita}}, \bibinfo {author} {\bibfnamefont {J.}~\bibnamefont {Ishizuka}}, \bibinfo {author} {\bibfnamefont {A.}~\bibnamefont {Daido}},\ and\ \bibinfo {author} {\bibfnamefont {Y.}~\bibnamefont {Yanase}},\ }\bibfield  {title} {\bibinfo {title} {Superconductivity in monolayer fese enhanced by quantum geometry},\ }\bibfield  {journal} {\bibinfo  {journal} {Physical Review Research}\ }\textbf {\bibinfo {volume} {4}},\ \href {https://doi.org/10.1103/physrevresearch.4.023232} {10.1103/physrevresearch.4.023232} (\bibinfo {year} {2022}{\natexlab{b}})\BibitemShut {NoStop}%
\bibitem [{\citenamefont {Hu}\ and\ \citenamefont {Huang}(2025)}]{Hu2025a}%
  \BibitemOpen
  \bibfield  {author} {\bibinfo {author} {\bibfnamefont {Y.-J.}\ \bibnamefont {Hu}}\ and\ \bibinfo {author} {\bibfnamefont {W.}~\bibnamefont {Huang}},\ }\bibfield  {title} {\bibinfo {title} {Quantum geometric superfluid weight in multiband superconductors: A microscopic interpretation},\ }\bibfield  {journal} {\bibinfo  {journal} {Physical Review B}\ }\textbf {\bibinfo {volume} {111}},\ \href {https://doi.org/10.1103/physrevb.111.134511} {10.1103/physrevb.111.134511} (\bibinfo {year} {2025})\BibitemShut {NoStop}%
\bibitem [{\citenamefont {Bailin}\ and\ \citenamefont {Love}(1982)}]{Bailin1982}%
  \BibitemOpen
  \bibfield  {author} {\bibinfo {author} {\bibfnamefont {D.}~\bibnamefont {Bailin}}\ and\ \bibinfo {author} {\bibfnamefont {A.}~\bibnamefont {Love}},\ }\bibfield  {title} {\bibinfo {title} {Superconductivity for relativistic electrons},\ }\href {https://doi.org/10.1088/0305-4470/15/9/046} {\bibfield  {journal} {\bibinfo  {journal} {Journal of Physics A: Mathematical and General}\ }\textbf {\bibinfo {volume} {15}},\ \bibinfo {pages} {3001–3005} (\bibinfo {year} {1982})}\BibitemShut {NoStop}%
\bibitem [{\citenamefont {Capelle}\ and\ \citenamefont {Gross}(1995)}]{Capelle1995}%
  \BibitemOpen
  \bibfield  {author} {\bibinfo {author} {\bibfnamefont {K.}~\bibnamefont {Capelle}}\ and\ \bibinfo {author} {\bibfnamefont {E.}~\bibnamefont {Gross}},\ }\bibfield  {title} {\bibinfo {title} {Relativistic theory of superconductivity},\ }\href {https://doi.org/10.1016/0375-9601(94)01010-r} {\bibfield  {journal} {\bibinfo  {journal} {Physics Letters A}\ }\textbf {\bibinfo {volume} {198}},\ \bibinfo {pages} {261–266} (\bibinfo {year} {1995})}\BibitemShut {NoStop}%
\bibitem [{\citenamefont {Strange}(1998)}]{Strange1998}%
  \BibitemOpen
  \bibfield  {author} {\bibinfo {author} {\bibfnamefont {P.}~\bibnamefont {Strange}},\ }\href {https://doi.org/10.1017/cbo9780511622755} {\emph {\bibinfo {title} {Relativistic Quantum Mechanics: With Applications in Condensed Matter and Atomic Physics}}}\ (\bibinfo  {publisher} {Cambridge University Press},\ \bibinfo {year} {1998})\BibitemShut {NoStop}%
\bibitem [{\citenamefont {Capelle}(2001)}]{Capelle2001}%
  \BibitemOpen
  \bibfield  {author} {\bibinfo {author} {\bibfnamefont {K.}~\bibnamefont {Capelle}},\ }\bibfield  {title} {\bibinfo {title} {Relativistic fluctuations and anomalous darwin terms in superconductors},\ }\bibfield  {journal} {\bibinfo  {journal} {Physical Review B}\ }\textbf {\bibinfo {volume} {63}},\ \href {https://doi.org/10.1103/physrevb.63.052503} {10.1103/physrevb.63.052503} (\bibinfo {year} {2001})\BibitemShut {NoStop}%
\bibitem [{\citenamefont {Gosselin}\ and\ \citenamefont {Mohrbach}(2010)}]{Gosselin2010}%
  \BibitemOpen
  \bibfield  {author} {\bibinfo {author} {\bibfnamefont {P.}~\bibnamefont {Gosselin}}\ and\ \bibinfo {author} {\bibfnamefont {H.}~\bibnamefont {Mohrbach}},\ }\bibfield  {title} {\bibinfo {title} {Appearance of gauge fields and forces beyond the adiabatic approximation},\ }\href {https://doi.org/10.1088/1751-8113/43/35/354025} {\bibfield  {journal} {\bibinfo  {journal} {Journal of Physics A: Mathematical and Theoretical}\ }\textbf {\bibinfo {volume} {43}},\ \bibinfo {pages} {354025} (\bibinfo {year} {2010})}\BibitemShut {NoStop}%
\bibitem [{\citenamefont {T\"{o}rm\"{a}}\ \emph {et~al.}(2018)\citenamefont {T\"{o}rm\"{a}}, \citenamefont {Liang},\ and\ \citenamefont {Peotta}}]{Törmä2018}%
  \BibitemOpen
  \bibfield  {author} {\bibinfo {author} {\bibfnamefont {P.}~\bibnamefont {T\"{o}rm\"{a}}}, \bibinfo {author} {\bibfnamefont {L.}~\bibnamefont {Liang}},\ and\ \bibinfo {author} {\bibfnamefont {S.}~\bibnamefont {Peotta}},\ }\bibfield  {title} {\bibinfo {title} {Quantum metric and effective mass of a two-body bound state in a flat band},\ }\bibfield  {journal} {\bibinfo  {journal} {Physical Review B}\ }\textbf {\bibinfo {volume} {98}},\ \href {https://doi.org/10.1103/physrevb.98.220511} {10.1103/physrevb.98.220511} (\bibinfo {year} {2018})\BibitemShut {NoStop}%
\bibitem [{\citenamefont {Iskin}(2022)}]{Iskin2022}%
  \BibitemOpen
  \bibfield  {author} {\bibinfo {author} {\bibfnamefont {M.}~\bibnamefont {Iskin}},\ }\bibfield  {title} {\bibinfo {title} {Effective-mass tensor of the two-body bound states and the quantum-metric tensor of the underlying bloch states in multiband lattices},\ }\bibfield  {journal} {\bibinfo  {journal} {Physical Review A}\ }\textbf {\bibinfo {volume} {105}},\ \href {https://doi.org/10.1103/physreva.105.023312} {10.1103/physreva.105.023312} (\bibinfo {year} {2022})\BibitemShut {NoStop}%
\bibitem [{\citenamefont {Iskin}(2024)}]{Iskin2024}%
  \BibitemOpen
  \bibfield  {author} {\bibinfo {author} {\bibfnamefont {M.}~\bibnamefont {Iskin}},\ }\bibfield  {title} {\bibinfo {title} {Cooper pairing, flat-band superconductivity, and quantum geometry in the pyrochlore-hubbard model},\ }\bibfield  {journal} {\bibinfo  {journal} {Physical Review B}\ }\textbf {\bibinfo {volume} {109}},\ \href {https://doi.org/10.1103/physrevb.109.174508} {10.1103/physrevb.109.174508} (\bibinfo {year} {2024})\BibitemShut {NoStop}%
\bibitem [{\citenamefont {Iskin}(2025)}]{Iskin2025}%
  \BibitemOpen
  \bibfield  {author} {\bibinfo {author} {\bibfnamefont {M.}~\bibnamefont {Iskin}},\ }\bibfield  {title} {\bibinfo {title} {Pair size and quantum geometry in a multiband hubbard model},\ }\bibfield  {journal} {\bibinfo  {journal} {Physical Review B}\ }\textbf {\bibinfo {volume} {111}},\ \href {https://doi.org/10.1103/physrevb.111.014502} {10.1103/physrevb.111.014502} (\bibinfo {year} {2025})\BibitemShut {NoStop}%
\bibitem [{\citenamefont {Tinkham}(2004)}]{Tinkham2004}%
  \BibitemOpen
  \bibfield  {author} {\bibinfo {author} {\bibfnamefont {M.}~\bibnamefont {Tinkham}},\ }\href@noop {} {\emph {\bibinfo {title} {Introduction to Superconductivity}}},\ \bibinfo {edition} {2nd}\ ed.\ (\bibinfo  {publisher} {Dover Publications},\ \bibinfo {year} {2004})\BibitemShut {NoStop}%
\bibitem [{\citenamefont {Cooper}(1956)}]{Cooper1956}%
  \BibitemOpen
  \bibfield  {author} {\bibinfo {author} {\bibfnamefont {L.~N.}\ \bibnamefont {Cooper}},\ }\bibfield  {title} {\bibinfo {title} {Bound electron pairs in a degenerate fermi gas},\ }\href@noop {} {\bibfield  {journal} {\bibinfo  {journal} {Physical Review}\ }\textbf {\bibinfo {volume} {104}} (\bibinfo {year} {1956})}\BibitemShut {NoStop}%
\bibitem [{\citenamefont {Bardeen}\ \emph {et~al.}(1957)\citenamefont {Bardeen}, \citenamefont {Cooper},\ and\ \citenamefont {Schrieffer}}]{BCS57}%
  \BibitemOpen
  \bibfield  {author} {\bibinfo {author} {\bibfnamefont {J.}~\bibnamefont {Bardeen}}, \bibinfo {author} {\bibfnamefont {L.~N.}\ \bibnamefont {Cooper}},\ and\ \bibinfo {author} {\bibfnamefont {J.~R.}\ \bibnamefont {Schrieffer}},\ }\bibfield  {title} {\bibinfo {title} {Theory of superconductivity},\ }\href {https://doi.org/10.1103/PhysRev.108.1175} {\bibfield  {journal} {\bibinfo  {journal} {Phys. Rev.}\ }\textbf {\bibinfo {volume} {108}},\ \bibinfo {pages} {1175} (\bibinfo {year} {1957})}\BibitemShut {NoStop}%
\bibitem [{\citenamefont {Puri}(2001)}]{Puri2001}%
  \BibitemOpen
  \bibfield  {author} {\bibinfo {author} {\bibfnamefont {R.~R.}\ \bibnamefont {Puri}},\ }\href {https://doi.org/10.1007/978-3-540-44953-9} {\emph {\bibinfo {title} {Mathematical Methods of Quantum Optics}}}\ (\bibinfo  {publisher} {Springer Berlin Heidelberg},\ \bibinfo {year} {2001})\BibitemShut {NoStop}%
\bibitem [{\citenamefont {Hall}(2015)}]{Hall2015}%
  \BibitemOpen
  \bibfield  {author} {\bibinfo {author} {\bibfnamefont {B.~C.}\ \bibnamefont {Hall}},\ }\href {https://doi.org/10.1007/978-3-319-13467-3} {\emph {\bibinfo {title} {Lie Groups, Lie Algebras, and Representations: An Elementary Introduction}}}\ (\bibinfo  {publisher} {Springer International Publishing},\ \bibinfo {year} {2015})\BibitemShut {NoStop}%
\bibitem [{\citenamefont {Hasan}\ and\ \citenamefont {Kane}(2010)}]{Hasan2010}%
  \BibitemOpen
  \bibfield  {author} {\bibinfo {author} {\bibfnamefont {M.~Z.}\ \bibnamefont {Hasan}}\ and\ \bibinfo {author} {\bibfnamefont {C.~L.}\ \bibnamefont {Kane}},\ }\bibfield  {title} {\bibinfo {title} {Colloquium: Topological insulators},\ }\href {https://doi.org/10.1103/revmodphys.82.3045} {\bibfield  {journal} {\bibinfo  {journal} {Reviews of Modern Physics}\ }\textbf {\bibinfo {volume} {82}},\ \bibinfo {pages} {3045–3067} (\bibinfo {year} {2010})}\BibitemShut {NoStop}%
\bibitem [{\citenamefont {Graf}\ and\ \citenamefont {Piéchon}(2021)}]{Graf2021}%
  \BibitemOpen
  \bibfield  {author} {\bibinfo {author} {\bibfnamefont {A.}~\bibnamefont {Graf}}\ and\ \bibinfo {author} {\bibfnamefont {F.}~\bibnamefont {Piéchon}},\ }\bibfield  {title} {\bibinfo {title} {Berry curvature and quantum metric in $n$-band systems: An eigenprojector approach},\ }\bibfield  {journal} {\bibinfo  {journal} {Physical Review B}\ }\textbf {\bibinfo {volume} {104}},\ \href {https://doi.org/10.1103/physrevb.104.085114} {10.1103/physrevb.104.085114} (\bibinfo {year} {2021})\BibitemShut {NoStop}%
\end{thebibliography}%
\appendix
\section{Derivation of Eqs.~(\ref{eq:QGT-Covpos},\ref{eq:Generator-position})}
\label{Appendix:QGT-pos}
From $\ket{u_n(\bm{k})}=U(\bm{k})\ket{\psi_n(\bm{k})}$ with $U(\bm{k})=e^{-i\bm{k}\cdot\hat{\bm{r}}}$, the derivative of the cell-periodic Bloch state reads
\begin{equation}
    \ket{\partial_{\mu}u_n}=\partial_{\mu}U\ket{\psi_n}+U\ket{\partial_{\mu}\psi_n}.
\end{equation}
Plugging the latter in the QGT yields
\begin{widetext}
\begin{subequations}
\begin{align}
        Q^{n}_{\mu\nu}&=\bra{\partial_{\mu}u_n}\big(\mathbbm{1}-\ket{u_n}\bra{u_n}\big)\ket{\partial_{\nu}u_n}=\big(\bra{\psi_n}\partial_{\mu}U^{\dagger}+\bra{\partial_{\mu}\psi_n}U^{\dagger}\big)\big(\mathbbm{1}-U\ket{\psi_n}\bra{\psi_n}U^{\dagger}\big)\big(\partial_{\nu}U\ket{\psi_n}+U\ket{\partial_{\nu}\psi_n}\big)\\
        &=\big(\bra{\psi_n}\partial_{\mu}U^{\dagger}U+\bra{\partial_{\mu}\psi_n}\big)\big(\mathbbm{1}-\ket{\psi_n}\bra{\psi_n}\big)\big(U^{\dagger}\partial_{\nu}U\ket{\psi_n}+\ket{\partial_{\nu}\psi_n}\big)\\
        &=\bra{\psi_n}\partial_{\mu}U^{\dagger}U\big(\mathbbm{1}-\ket{\psi_n}\bra{\psi_n}\big)U^{\dagger}\partial_{\nu}U\ket{\psi_n}+\sum_{m\neq n}\big(\bra{\psi_n}\partial_{\mu}U^{\dagger}U\ket{\psi_m}\bra{\psi_m}\ket{\partial_{\nu}\psi_n}+\bra{\partial_{\mu}\psi_n}\ket{\psi_m}\bra{\psi_m}U^{\dagger}\partial_{\nu}U\ket{\psi_n}\nonumber\\
        &+\bra{\partial_{\mu}\psi_n}\ket{\psi_m}\bra{\psi_m}\ket{\partial_{\nu}\psi_n}\big).
\end{align}
\end{subequations}
\end{widetext}
Since the Bloch states $\ket{\psi_n(\bm{k})}$ are the eigenstates of the $\bm{k}$-independent Hamiltonian $H$, they form an orthonormal basis by the spectral theorem \cite{Cayssol2021}. Using this, for $m\neq n$ we have
\begin{align}
    \bra{\psi_m(\bm{k})}\ket{\partial_{\mu}\psi_n(\bm{k})}&=\lim_{h\to0}\frac{1}{h}\big[\bra{\psi_m(\bm{k})}\ket{\psi_n(\bm{k}+h\bm{e}_{\mu})}\nonumber\\
    -\bra{\psi_m(\bm{k})}&\ket{\psi_n(\bm{k})}]=0.
\end{align}
From $(\partial_{\mu}U^{\dagger})U=-U^{\dagger}\partial_{\mu}U$, the QGT then reads
\begin{subequations}
\begin{align}
    Q^{n}_{\mu\nu}&=-\bra{\psi_n}\partial_{\mu}U^{\dagger}U\big(\mathbbm{1}-\ket{\psi_n}\bra{\psi_n}\big)\partial_{\nu}U^{\dagger}U\ket{\psi_n}\\
    &=-\Cov_{\ket{\psi_n}}\big(\partial_{\mu}U^{\dagger}U,\partial_{\nu}U^{\dagger}U\big)\\
    &=\Cov_{\ket{\psi_n}}\big(-i\partial_{\mu}U^{\dagger}U,-i\partial_{\nu}U^{\dagger}U\big)\\
    &=\Cov_{\ket{\psi_n}}\big(\hat{R}_{\mu},\hat{R}_{\nu}\big).
\end{align}
\end{subequations}
The generators $\hat{R}_{\mu}$ are given by
\begin{equation}
    \hat{R}_{\mu}=-i\partial_{\mu}U^{\dagger}U=-i(\partial_{\mu}e^{i\bm{k}\cdot\hat{\bm{r}}})e^{-i\bm{k}\cdot\hat{\bm{r}}}.
\end{equation}
An operator $\hat{A}$ depending on a variable $\lambda\in\mathbb{R}$ obeys  the following identity \cite{Liu2019,Puri2001},
\begin{subequations}
\begin{align}
    \partial_{\lambda}e^{\hat{A}}&=\int_{0}^{1}e^{s\hat{A}}\partial_{\lambda}\hat{A}e^{(1-s)\hat{A}}\dd s\\
    \Leftrightarrow\hspace{1mm}\partial_{\lambda}e^{\hat{A}}e^{-\hat{A}}&=\int_{0}^{1}e^{s\hat{A}}\partial_{\lambda}\hat{A}e^{-s\hat{A}}\dd s.
\end{align}
\end{subequations}
For $\hat{A}=i\bm{k}\cdot\hat{\bm{r}}$ and $\lambda=k_{\mu}$, this yields
\begin{subequations}
\begin{align}
    (\partial_{\mu}e^{i\bm{k}\cdot\hat{\bm{r}}})e^{-i\bm{k}\cdot\hat{\bm{r}}}&=\int_{0}^{1}e^{is\bm{k}\cdot\hat{\bm{r}}}\partial_{\mu}(i\bm{k}\cdot\hat{\bm{r}})e^{-is\bm{k}\cdot\hat{\bm{r}}}\dd s\\
    &=i\int_{0}^{1}e^{is\bm{k}\cdot\hat{\bm{r}}}\hat{r}_{\mu}e^{-is\bm{k}\cdot\hat{\bm{r}}}\dd s.
\end{align}
\end{subequations}
The generator $\hat{R}_{\mu}$ is then readily expressed as follows,
\begin{equation}
    \hat{R}_{\mu}=\int_{0}^{1}e^{is\bm{k}\cdot\hat{\bm{r}}}\hat{r}_{\mu}e^{-is\bm{k}\cdot\hat{\bm{r}}}\dd s.
\end{equation}
Using the identity $e^{X}Ye^{-X}=e^{\ad_{X}}Y$ \cite{Hall2015} and the linearity of the adjoint representation, we further obtain
\begin{subequations}
    \begin{align}
        \hat{R}_{\mu}&=\int_{0}^{1}e^{is\ad_{\bm{k}\cdot\hat{\bm{r}}}}\hat{r}_{\mu}\dd s=\Bigg(\int_{0}^{1}e^{is\ad_{\bm{k}\cdot\hat{\bm{r}}}}\dd s\Bigg) \hat{r}_{\mu}\\
        &=\Bigg(\sum_{n=0}^{+\infty}\frac{i^n\ad^n_{\bm{k}\cdot\hat{\bm{r}}}}{n!}\int_{0}^{1}s^{n}\dd s\Bigg)\hat{r}_{\mu}\\
        &=\sum_{n=0}^{+\infty}\frac{i^n}{(n+1)!}\ad^{n}_{\bm{k}\cdot\hat{\bm{r}}}\hat{r}_{\mu}.
    \end{align}
\end{subequations}

\section{2d massive Dirac fermions}
\label{annexe:Massive-Dirac}
The low-energy Bloch Hamiltonian describing two-dimensional massive Dirac fermions is given by \cite{Hasan2010}
\begin{equation}
    H(\bm{k})=\hbar vk_x\sigma_x+\hbar vk_y\sigma_y+\Delta\sigma_z,
\end{equation}
with $\sigma_{\mu}$ the Pauli matrices, $v$ the Dirac velocity and $\Delta$ the band gap. The band dispersions are $\epsilon_{n}=n|\Delta|\sqrt{1+\lambdabar_c^2k^2}$, with $n=\pm1$ and $\lambdabar_c=\hbar v/|\Delta|$ the reduced Compton wavelength. Quantum geometry is readily computed by means of the Bloch vector $\bm{b}_n=n\bm{h}/|\bm{h}|$, with $H=\bm{h}\cdot\bm{\sigma}$ \cite{Graf2021},
\begin{subequations}
    \begin{align}
        g^{n}_{\mu\nu}&=\frac{1}{2}\partial_{\mu}\bm{b}_{n}\cdot\partial_{\nu}\bm{b}_{n},\\
        \mathcal{B}^{n}_{\mu\nu}&=-\frac{1}{2}\bm{b}_{n}\cdot\partial_{\mu}\bm{b}_{n}\times\partial_{\nu}\bm{b}_n.
    \end{align}
\end{subequations}
The Bloch vector is given by 
\begin{equation}
    \bm{b}_{n}=\frac{1}{\epsilon_n}\big(\hbar vk_x,\hbar vk_y,\Delta\big),
\end{equation}
which directly yields
\begin{subequations}
    \label{eq:QGeom-MDirac}
    \begin{align}
        g^{n}_{\mu\nu}&=\frac{\lambdabar_c^2}{4}\frac{(1+\lambdabar_c^2k^2)\delta_{\mu\nu}-\lambdabar_c^2k_{\mu}k_{\nu}}{(1+\lambdabar_c^2k^2)^2},\\
        \mathcal{B}^{n}_{xy}&=-\frac{n\sgn\Delta}{2}\frac{\lambdabar_c^2}{(1+\lambdabar_c^2k^2)^{3/2}}.
    \end{align}
\end{subequations}
At $k=0$, i.e. at the $K$ point, we have $g^{n}_{xy}=0$ and $g^{n}_{xx}=g^{n}_{yy}=\lambdabar_c^2/4$. Additionally, $|\mathcal{B}^{n}_{xy}|=\lambdabar_c^2/2$. At the $K$ point we therefore have
\begin{subequations}
\begin{align}
    \Tr g^{n}&=\frac{\lambdabar_c^2}{2}=|\mathcal{B}^{n}_{xy}|,\\
    \sqrt{\det g^{n}}&=\sqrt{g^{n}_{xx}g^{n}_{yy}}=\frac{\lambdabar_c^2}{4}=\frac{1}{2}|\mathcal{B}^{n}_{xy}|.
\end{align}
\end{subequations}

\section{Derivation of Eq.~(\ref{eq:Fluct-2b-QG})}
\label{annexe:2bodypb}
The fluctuation of $\hat{\bm{\rho}}$ is expressed as
\begin{equation}
    \delta\hat{\rho}_{\mu}=\hat{\rho}_{\mu}-\langle\hat{\rho}\rangle_{\mu}=\delta\hat{r}_{1\mu}-\delta\hat{r}_{2\mu},
\end{equation}
with $\delta\hat{r}_{1\mu}=\hat{r}_{1\mu}-\langle\hat{r}_{1\mu}\rangle_{n}$ and $\delta\hat{r}_{2\mu}=\hat{r}_{2\mu}-\langle\hat{r}_{2\mu}\rangle_{m}$
The second order fluctuations $\langle\delta\hat{\rho}_{\mu}\delta\hat{\rho}_{\nu}\rangle_{\Psi}$ is expanded as follows,
\begin{align}
    \langle\delta\hat{\rho}_{\mu}&\delta\hat{\rho}_{\nu}\rangle_{\Psi}=\langle\delta\hat{r}_{1\mu}\delta\hat{r}_{1\nu}\rangle_{\Psi}-\langle\delta\hat{r}_{1\mu}\delta\hat{r}_{2\nu}\rangle_{\Psi}\nonumber\\
    &-\langle\delta\hat{r}_{2\mu}\delta\hat{r}_{1\nu}\rangle_{\Psi}+\langle\delta\hat{r}_{2\mu}\delta\hat{r}_{2\nu}\rangle_{\Psi}.
\end{align}
Using the identity $(A\otimes B)(C\otimes D)=AC\otimes BD$, the cross-terms can be shown to vanish as follows
\begin{widetext}
\begin{subequations}
    \begin{align}
        \langle\delta\hat{r}_{1\mu}\delta\hat{r}_{2\nu}\rangle_{\Psi}&=\bra{\psi_n}\bra{\psi_m}\big(\hat{r}_{1\mu}\otimes\mathbbm{1}-\langle\hat{r}_{1\mu}\rangle_{n}\big)\big(\mathbbm{1}\otimes\hat{r}_{2\nu}-\langle\hat{r}_{2\nu}\rangle_{m}\big)\ket{\psi_n}\ket{\psi_m}\\
        &=\bra{\psi_n}\bra{\psi_m}\big(\hat{r}_{1\mu}-\langle\hat{r}_{1\mu}\rangle_{n}\big)\otimes\big(\hat{r}_{2\nu}-\langle\hat{r}_{2\nu}\rangle_{m}\big)\ket{\psi_n}\ket{\psi_m}\\
        &=\langle\hat{r}_{1\mu}-\langle\hat{r}_{1\mu}\rangle_{n}\rangle_{n}\langle\hat{r}_{2\nu}-\langle\hat{r}_{2\nu}\rangle_{m}\rangle_{m}=0,
    \end{align}
\end{subequations}
\end{widetext}
where for brevity we have denoted $\ket{\psi_n}\ket{\psi_m}=\ket{\psi_n}\otimes\ket{\psi_m}$. Physically, this means that since the Hilbert space of interest is a tensor product of the Hilbert spaces of the two Bloch fermions, the quantum fluctuations of their respective position are uncorrelated and as such $\langle\delta\hat{r}_{1\mu}\delta\hat{r}_{2\nu}\rangle_{\Psi}=\langle\delta\hat{r}_{1\mu}\rangle_{\Psi}\langle\delta\hat{r}_{2\nu}\rangle_{\Psi}=0$. That way, the two cross terms disappear and we have 
\begin{equation}
    \langle\delta\hat{\rho}_{1\mu}\delta\hat{\rho}_{2\nu}\rangle_{\Psi}=\langle\delta\hat{r}_{1\mu}\delta\hat{r}_{1\nu}\rangle_{\Psi}+\langle\delta\hat{r}_{2\mu}\delta\hat{r}_{2\nu}\rangle_{\Psi}.
\end{equation}
We further have
\begin{subequations}
    \begin{align}
        \langle\delta\hat{r}_{1\mu}\delta\hat{r}_{1\mu}\rangle_{\Psi}&=\bra{\psi_n}\bra{\psi_m}\delta\hat{r}_{1\mu}\delta\hat{r}_{1\nu}\otimes\mathbbm{1}\ket{\psi_n}\ket{\psi_m}\\
        &=\langle\delta\hat{r}_{1\mu}\delta\hat{r}_{1\nu}\rangle_{n}=Q^{n}_{\mu\nu}.
    \end{align}
\end{subequations}
Similarly, $\langle\delta\hat{r}_{2\mu}\delta\hat{r}_{2\nu}\rangle_{\Psi}=Q^{m}_{\mu\nu}$. We thus finally obtain
\begin{equation}
    \langle\delta\hat{\rho}_{\mu}\delta\hat{\rho}_{\nu}\rangle_{\Psi}=Q^{n}_{\mu\nu}(\bm{k}_{1})+Q^{m}_{\mu\nu}(\bm{k}_{2}),
\end{equation}
where we have reintroduced the momentum dependences.

\section{Calculation of $\langle V^{\text{eff}}_{\bm{k}\bm{k'}}\rangle$ for 2d massive Dirac fermions}
\label{annexe:Cooper}
From Eq.~(\ref{eq:QGeom-MDirac}), the elements of the quantum metric for the conduction band $n=+1$ are given by
\begin{equation}
    g^{n}_{xx}=\frac{\lambdabar_c^2}{4}\frac{1+\lambdabar_c^2k^2_y}{(1+\lambdabar_c^2k^2)^2},\hspace{2mm}g^{n}_{yy}=\frac{\lambdabar_c^2}{4}\frac{1+\lambdabar_c^2k^2_x}{(1+\lambdabar_c^2k^2)^2},
\end{equation}
and 
\begin{equation}
    g^{n}_{xy}=-\frac{\lambdabar_c^2}{4}\frac{\lambdabar_c^2k_xk_y}{(1+\lambdabar_c^2k^2)^{2}}.
\end{equation}
We therefore have
\begin{align}
    g^{n}_{\mu\nu}k_{\mu}k_{\nu}&=\frac{\lambdabar_c^2}{4(1+\lambdabar_c^2k^2)^2}\big[(1+\lambdabar_c^2k^2_y)k_x^2+(1+\lambdabar_c^2k_x^2)k_y^2\nonumber\\
    &-2\lambdabar_c^2k_x^2k_y^2\big]=\frac{\lambdabar_c^2k^2}{4(1+\lambdabar_c^2k^2)^2}.
\end{align}
For the quadratic average $\langle k_{\mu}'k_{\nu}'\rangle$, the symmetry of the set $\mathcal{D}$ can be invoked to show that $\langle k_{x}'k_{y}'\rangle$ vanishes. We thus have
\begin{subequations}
    \begin{align}
        &g_{\mu\nu}^{n}\langle k_{\mu}'k_{\nu}'\rangle=\frac{\lambdabar_c^2}{4(1+\lambdabar_c^2k^2)^2}\big[(1+\lambdabar_c^2k_y^2)\langle k'_{x}{}^{2}\rangle\nonumber\\
        &+(1+\lambdabar_c^2k_x^2)\langle k'_{y}{}^{2}\rangle\big]\\
        &=\frac{\lambdabar_c^2\langle k'{}^{2}\rangle+\lambdabar_c^4(k_x^2\langle k_{y}'{}^{2}\rangle+k_{y}^2\langle k_{x}'{}^{2}\rangle)}{4(1+\lambdabar_c^2k^2)^2}.
    \end{align}
\end{subequations}
Supposing that the presence of quantum geometry doesn't introduce anisotropy in the Cooper pair's wavefunction, for $s$-wave pairing the coefficients $p_{\bm{k}}$ depend only on the modulus of $\bm{k}$. This isotropy means that $\langle k_x'{}^2\rangle=\langle k_y'{}^{2}\rangle=\langle k'{}^{2}\rangle/2$, such that
\begin{subequations}
    \begin{align}
        g_{\mu\nu}^{n}\langle k_{\mu}'k_{\nu}'\rangle&=\frac{\lambdabar_c^2\langle k'{}^{2}\rangle+\lambdabar_c^4k^2\langle k'{}^{2}\rangle/2}{4(1+\lambdabar_c^2k^2)^2}\\
        &=\frac{\lambdabar_c^2\langle k'{}^{2}\rangle(1+\lambdabar_c^2k^2/2)}{4(1+\lambdabar_c^2k^2)^2}.
    \end{align}
\end{subequations}
We thus obtain 
\begin{equation}
    \frac{\langle V^{\text{eff}}_{\bm{k}\bm{k'}}\rangle}{-V}=1-\frac{\lambdabar_c^2k^2+\lambdabar_c^2\langle k'{}^{2}\rangle(1+\lambdabar_c^2k^2/2)}{4(1+\lambdabar_c^2k^2)^2}.
\end{equation}

\end{document}